\documentclass[12pt]{article}
\usepackage[square,numbers,sort&compress]{natbib}
\usepackage{enumitem}
\usepackage{amsfonts}
\usepackage{amsmath}
\usepackage{graphicx}
\usepackage{subfigure}
\usepackage{fancyhdr}
\usepackage{amssymb}
\usepackage{float}
\usepackage{verbatim}
\usepackage{algorithm, algorithmic}
\usepackage{url} 
\usepackage[square,numbers,sort&compress]{natbib}
\usepackage{array,graphicx}
\usepackage{enumerate}
\usepackage{graphicx}
\usepackage{multirow}
\usepackage{subfigure}
\usepackage{leftidx}
\usepackage{setspace}
\usepackage{tikz}
\usetikzlibrary{shapes.geometric, arrows}
\usepackage[titletoc,toc,title]{appendix}
\usepackage{amsmath}
\usepackage{graphicx}

\usepackage{colortbl}
\usepackage{booktabs}
\usepackage{adjustbox}
\usepackage{relsize}
\usepackage{booktabs}
\usepackage{diagbox}
\usepackage{extarrows}
\usepackage{multirow}

\usepackage{setspace}
\makeatletter
\newenvironment{breakablealgorithm}
  {
   \begin{center}
     \refstepcounter{algorithm}
     \hrule height.8pt depth0pt \kern2pt
     \renewcommand{\caption}[2][\relax]{
       {\raggedright\textbf{\ALG@name~\thealgorithm} ##2\par}%
       \ifx\relax##1\relax 
         \addcontentsline{loa}{algorithm}{\protect\numberline{\thealgorithm}##2}%
       \else 
         \addcontentsline{loa}{algorithm}{\protect\numberline{\thealgorithm}##1}%
       \fi
       \kern2pt\hrule\kern2pt
     }
  }{
     \kern2pt\hrule\relax
   \end{center}
  }
\makeatother

\begin{document}
\title{Pricing American options with exogenous and endogenous transaction costs}
\author{{Dong Yan$^{a}$, Xin-Jie Huang$^{a}$, Guiyuan Ma$^{b}$, Xin-Jiang He$^{c,d,}$\thanks{Corresponding author: xinjiang@zjut.edu.cn.}}\\
\small\it a. School of Statistics, University of International Business and Economics, Beijing, China.\\
\small\it b. School of Economics and Finance, Xi'an Jiaotong University, Xi'an, China.\\
{\small\it c. School of Economics, Zhejiang University of Technology, Hangzhou, China.}\\
{\small\it d. Institute for Industrial System Modernization, Zhejiang University of Technology, Hangzhou, China.}
}
\date{}
\maketitle

\begin{abstract}
We study an American option pricing problem with liquidity risks and transaction fees. As endogenous transaction costs, liquidity risks of the underlying asset are modeled by a mean-reverting process. Transaction fees are exogenous transaction costs and are assumed to be proportional to the trading amount, with the long-run liquidity level depending on the proportional transaction costs rate. Two nonlinear partial differential equations are established to characterize the option values for the holder and the writer, respectively. To illustrate the impact of these transaction costs on option prices and optimal exercise prices, we apply the alternating direction implicit method to solve the linear complementarity problem numerically. Finally, we conduct model calibration from market data via maximum likelihood estimation, and find that our model incorporating liquidity risks outperforms the Leland model significantly.

\end{abstract}


\section{Introduction}
Black and Scholes \cite{black1973pricing} first provided a closed-form pricing formula for the European call option, which laid the solid foundation for the modern theory of pricing financial derivatives.  As we know, the valuation of American options remains one of the most challenging problems in derivative pricing due to its nonlinearity, and there is still no closed-form solution. Brennan and Schwartz \cite{brennan1977valuation} proposed an algorithm to price American put options. Cox et al. \cite{cox1979option} presented a simple binomial tree method for American option pricing. Carr et al. \cite{carr1992alternative} derived alternative representations of the McKean equation for pricing American put options. Longstaff \cite{longstaff2001valuing} provided a simple least-squares approach for approximating the value of American options via simulation. Since then, more and more advanced numerical schemes have been developed to price American options by incorporating more realistic factors, such as homotopy perturbation methods \cite{zhu2006exact,dehghan2011solution}, hard-to-borrow stocks \cite{ma2018pricing,ma2019pricing}, near expiry for dividend-paying stocks \cite{dehghan2017asymptotic}, as well as price jumps and economic regime shifts \cite{shirzadi2020pricing,shirzadi2023american}.

However,  there are some strong assumptions in the classic option pricing theory \cite{he2024analytically, he2024analytical2}. The most typical example is that it assumes the market is perfectly frictionless, which means that there are no costs associated with asset trading. However, transaction costs are in fact a major factor that needs to be taken into consideration when trading assets, and they can be categorized into two main types, i.e., exogenous transaction costs and endogenous transaction costs.

Exogenous transaction costs, which are also called transaction fees, generally include stamp duty in real market. While some paid attention to fixed transaction fees \cite{zakamouline2006european, sun2007reset}, many others focused on proportional transaction fees, which are closer to practice. Leland \cite{leland1985option} pioneeringly proposed an option pricing model with proportional transaction fees, and then developed a modified strategy via option replicating, where a nonlinear partial differential equation (PDE) was solved by Amster et al. \cite{amster2005black}. Davis et al. \cite{davis1993european} applied the utility indifference method to solve the problem numerically after incorporating proportional transaction fees. Their framework was further extended by Monoyios \cite{monoyios2004option} to price European options with a Markov chain approximation. A further extension was considered by Xing et al. \cite{xing2017european}, who developed a new algorithm to price European options under a geometric Levy process. Sevcovic and Zitnanska \cite{vsevvcovivc2016analysis} derived the option pricing formula with variable transaction fees and transformed the nonlinear equation into a quasilinear parabolic equation. Yan et al. \cite{yan2021utility} used a utility indifference method to price European options with proportional transaction fees, which innovatively considered the variability about the fraction of one's total wealth in the risky asset. Recently, Yan et al. \cite{yan2022pricing} introduced the proportional transaction fees for American option pricing with stochastic volatility.

On the other hand, liquidity cost, resulting from liquidity risks, is one type of endogenous costs that cannot be neglected, as it is almost ubiquitous in real markets. Much literature has explored about liquidity costs when pricing different assets, although there is no consensus on how to effectively model liquidity risks. For example, many new capital asset pricing models have been proposed to calculate asset returns when liquidity risks and/or illiquidity premiums were taken into account \cite{acharya2005asset, watanabe2008time, lee2011world}, while another bunch of researchers depicted liquidity risks with the bid-ask spread \cite{amihud1989effects, guillaume2019implied}. Very recently, there are also many new dynamic portfolio choice problems in which stochastic liquidity is incorporated \cite{collin2016insider,fruth2019optimal,ma2020optimal,ma2023dynamic,han2023strategic}.

Considering that option prices heavily depend on the corresponding underlying asset, there have already been a few works investigating the impact of underlying assets' liquidity levels on option prices. Market liquidity is a concept that was prevailing in this particular area, and it treats the market as the only counterparty to all transactions, so that the liquidity levels of all assets are affected by market liquidity. Madan and Cherny \cite{madan2010markets} modeled market liquidity through a constant parameter measuring market stress when pricing financial derivatives, which was then shown to be mean reverting and display a term structure \cite{albrecher2013implied}.  This is the motivation why Feng et al. \cite{feng2014option} modeled market liquidity as a stochastic process with mean-reversion (the Feng model). They followed the idea of Brunetti and Caldarera \cite{brunetti2004asset} by using a liquidity-dependent discount factor to obtain reduced underlying prices due to liquidity risks. Feng et al. \cite{feng2016importance} further empirically compared the results of their newly derived European option pricing formula with the traditional European option pricing formula, and verified that the new pricing formula presented smaller pricing errors. Similarly, European quanto options and discrete barrier options were priced in the presence of liquidity risk \cite{li2018european, li2019pricing}, while Zhang et al. \cite{zhang2019derivatives} considered the valuation of various financial derivatives. Pasricha et al. \cite{pasricha2022closed} generalized the Feng model by incorporating  a more general correlation structure among different Brownian motions, while maintaining the analytical tractability of the original model. More recently, He et al. \cite{he2025closed2, he2025closed3} moved a step further to combine stochastic volatility, regime switching and stochastic liquidity together when pricing European and exchange options.

In this paper, in order to study the impacts of both exogenous and endogenous transaction costs, we consider the valuation of American options when the underlying asset is affected by liquidity risks and its trading is subject to transaction costs. Following existing works \cite{feng2014option, he2025closed, lin2024closed}, we adopt a mean-reverting Ornstein-Uhlenbeck process to model stochastic liquidity. To establish the intrinsic connection between exogenous and endogenous transaction costs, we further assume that the long-run level of market liquidity is affected by transaction costs. With dynamic hedging, we employ a known option in the construction of the portfolio to hedge against the risk resourced from stochastic liquidity risk, and construct systems of PDEs for the holder and the writer of an American option, respectively, under the assumption of proportional transaction costs. To numerically solve the established highly nonlinear and high-dimensional PDE systems, we turn to the finite difference method, such as explicit, fully implicit, and Crank-Nicolson schemes. However, it will come across the curve of dimensionality when applying the explicit method. In other words, the explicit method suffers from severe stability constraints which are often too restrictive for practical use, while the fully implicit and Crank-Nicolson schemes generate large, computationally expensive systems of linear equations that are difficult to solve efficiently. To circumvent these issues, we adopt the alternating direction implicit (ADI) method. The ADI scheme is specifically designed for such multi-dimensional problems, offering an excellent combination of unconditional stability and computational speed by splitting the full dimensional problem into a series of one-dimensional calculations at each time step \cite{duffy2013finite}.

Our contributions in this manuscript lie in three aspects. From a modeling perspective, we develop a new American option pricing model that incorporates both liquidity risks and proportional transaction costs. From a mathematical standpoint, we apply the ADI method to solve the linear complementarity problem (LCP) to overcome the strong nonlinearity induced by transaction costs. From an empirical perspective, we calibrate the model using real-world data via maximum likelihood estimation (MLE) and demonstrate that our liquidity-adjusted model significantly outperforms the Leland model.

The remainder of this paper is organized as follows. We construct a model for pricing American options, which incorporates transaction costs and liquidity risks, and the pricing PDE systems are established in Section 2. We then design numerical schemes, and present some results of numerical experiments in Section 3. Section 4 uses market data to demonstrate that the revised model outperforms the benchmark model in option pricing,  with conclusions provided in the last section.

\section{Formulation of the model}

In this section, we proposed a new dynamic model for the underlying asset, where both liquidity risks and proportional transaction costs are taken into consideration. Similar to the literature \cite{pasricha2022closed, lin2025analytically}, the stock price $S$ and the liquidity risk $L$ are characterized as
\begin{equation}
\label{Dynamics_SL}
\begin{cases}
        d S_{t}=\mu S_t d t+\beta L_tS_t d W^\gamma_t+\sigma_SS_t d W^S_t,\\
         d L_{t}=\alpha(\theta-L_t)d t+\sigma_L d W^L_t,
\end{cases}
\end{equation}
where $W^\gamma$, $W^S$, $W^L$ are standard Brownian motions with the correlation structure specified as
\begin{equation}
\begin{cases}
   d W^\gamma_t d W^S_t=\rho_1dt,  \\
         dW_t^L d W^S_t=\rho_2 d t,\\
         d W^\gamma_t d W^L_t=\rho_3d t.
\end{cases}
\label{eqcorr}
\end{equation}
$\mu$ is the drift term of the risky asset and $\sigma_S$ is its constant volatility . The strictly positive parameter $\beta$ measures the sensitivity to the level of market liquidity of the asset price. The liquidity risk $L$ follows a mean-reverting process with the mean-reversion speed $\alpha$, the mean-reversion level $\theta$ and the volatility of market liquidity $\sigma_L$.

In the presence of transaction costs, three factors need to be considered in option pricing.

First, the assumption on perfect hedging under the conventional Black-Scholes framework is no longer possible, as hedging the portfolio continuously in this case would result in abnormally large trading costs. In order to ensure that limit the cumulated transaction costs during the life of the option, the investor will hedge the portfolio at discrete times.
   
Second, the option price is no longer unique for two parties since both holder and writer of an option wish trading costs can be compensated from the option premium. Instead, a fair price range with the holding price being the lower bound and the writing price being the upper bound should be given reasonably.

Last but not least, the effects of transaction costs on liquidity risks should not be neglected. Theoretically, an increase in the transaction costs rate will enlarge the market illiquidity level, since investors would be more reluctant to trade with the same amount of profit. To capture this, the long-run mean of the illiquidity level is assumed to be positively related to transaction costs rate. Specifically, we assume that the mean reversion level of liquidity risks in Eq. (\ref{Dynamics_SL}) is affected by the transaction costs rate, transaction costs sensitive coefficient $\lambda$ and the illiquidity level, i.e.,
\begin{equation}
\theta_t=\bar{\theta}+\kappa\cdot  \lambda \cdot g(L_t).
\end{equation}
Here, $\bar{\theta}$ is the mean reversion level without the effects of transaction costs as specified in Eq. (\ref{Dynamics_SL}). The function $g(L)$ should be chosen as a concave function because the impact of transaction costs on illiquidity decreases with increasing illiquidity levels. Specifically, when the illiquidity level is low, the effect of transaction costs grows more rapidly, whereas when the illiquidity level is already high, the marginal impact diminishes. Theoretically, any concave functional form with reasonable parameter values would be appropriate (e.g., $g(L)=log(1+\lambda L)$ with $\lambda>0$). For computational convenience, we adopt a power function specification, $g(L)=L^\zeta$ with $\zeta\in (0, 1)$, where different values of $\zeta$ allow us to modulate the function's convexity.

We now formulate the model for the option holding price. Consider a portfolio $\Pi$ consisting of one option $V (S, L, t)$ , $-\Delta$ of the underlying asset, and $-\Delta_1$ of a known option $V_1(S, L, t)$. The value of this portfolio is
\begin{equation}
\Pi=V(S,L,t)-\Delta S-\Delta_1V_1(S,L,t),
\end{equation}
where $V_1$ is chosen as the known option whose value satisfies the closed-form pricing formula in \cite{pasricha2022closed} without transaction costs.

Given that transaction cost accumulation prevents continuous hedging, we assume that the portfolio is hedged in a non-infinitesimal fixed time step $\delta t$\footnote{$\delta t=\frac{1}{12}$ corresponds to monthly hedging, while $\delta t =\frac{1}{4}$ represents quarterly hedging.}. Under the assumption of self-financing, the change in the value of the hedging portfolio during $\delta t$ is
\begin{equation*}
\delta\Pi=\delta V-\Delta\delta S-\Delta_{1}\delta V_{1}-\kappa S|\nu|,
\end{equation*}
where $\kappa$ is the proportional transaction costs rate, and $\nu$ presents the number of traded stocks during a non-infinitesimal fixed time-step.

Applying It$\hat{\text{o}}$'s lemma for $V$ and $V_1$ according to the dynamics of state variables in Eq. (\ref{Dynamics_SL}), we obtain
\begin{align*}
\delta \Pi&=\bigg[\frac{\partial V}{\partial t}+\frac{S^2}{2}\bigg(\beta^2L^2+\sigma_S^2+2\rho_1\sigma_S \beta L\bigg)\frac{\partial ^2 V}{\partial S^2}+\frac{1}{2}\sigma^2_L\frac{\partial ^2 V}{\partial L^2}+\bigg(\rho_3\sigma_L\beta L+\rho_2\sigma_L\sigma_S\bigg)S\frac{\partial ^2V}{\partial S\partial L}\bigg]\delta t\\
 &-\Delta_1 \bigg[\frac{\partial V_1}{\partial t}+\frac{S^2}{2}\bigg(\beta^2L^2+\sigma_S^2+2\rho_1\beta\sigma_S L\bigg)\frac{\partial ^2 V_1}{\partial S^2}+\frac{1}{2}\sigma^2_L\frac{\partial ^2 V_1}{\partial L^2}+\bigg(\rho_3\sigma_L\beta L+\rho_2\sigma_L\sigma_S\bigg)S\frac{\partial ^2V_1}{\partial S\partial L}\bigg]\delta t\\
 &+\bigg(\frac{\partial V}{\partial S}-\Delta_1\frac{\partial V_1}{\partial S}-\Delta\bigg)\delta S+\bigg(\frac{\partial V}{\partial L}-\Delta_1\frac{\partial V_1}{\partial L}\bigg)\delta L-\kappa S|\nu|.
\end{align*}
To hedge against the risks associated with the fluctuations of asset prices and liquidity risks, we choose
\begin{equation*}
\begin{cases}
\displaystyle        \frac{\partial V}{\partial S}-\Delta_1\frac{\partial V_1}{\partial S}-\Delta=0,\\
 \displaystyle         \frac{\partial V}{\partial L}-\Delta_1\frac{\partial V_1}{\partial L}=0,
\end{cases}
\end{equation*}
which gives
\begin{equation}
        \Delta=\frac{\partial V}{\partial S}-\Delta_{1}\frac{\partial V_1}{\partial S} \ \ \ \ \text{and}\ \ \ \ \Delta_1=\frac{\partial V}{\partial L}\cdot \bigg(\frac{\partial V_1}{\partial L}\bigg)^{-1}.
        \label{Delta&Delta_1}
\end{equation}
Rearranging the portfolio dynamics, we obtain
\begin{align}
\delta \Pi&=\bigg[\frac{\partial V}{\partial t}+\frac{S^2}{2}\bigg(\beta^2L^2+\sigma_S^2+2\rho_1\sigma_S \beta L\bigg)\frac{\partial ^2 V}{\partial S^2}+\frac{1}{2}\sigma^2_L\frac{\partial ^2 V}{\partial L^2}+\bigg(\rho_3\sigma_L\beta L+\rho_2\sigma_L\sigma_S\bigg)S\frac{\partial ^2V}{\partial S\partial L}\bigg]\delta t\nonumber\\
 &-\frac{\partial V}{\partial L} \cdot \frac{\bigg[\frac{\partial V_1}{\partial t}+\frac{S^2}{2}\bigg(\beta^2L^2+\sigma_S^2+2\rho_1\beta\sigma_S L\bigg)\frac{\partial ^2 V_1}{\partial S^2}+\frac{1}{2}\sigma^2_L\frac{\partial ^2 V_1}{\partial L^2}+\bigg(\rho_3\sigma_L\beta L+\rho_2\sigma_L\sigma_S\bigg)S\frac{\partial ^2V_1}{\partial S\partial L}\bigg]}{\frac{\partial V_1}{\partial L}}\delta t\nonumber\\
 &-\kappa S|\nu|.
 \label{deltaPi}
\end{align}
Under the no arbitrage argument, the investor may allocate the portfolio value $\Pi$ to a risk-free asset, yielding a growth of $r\Pi \delta t$ per time step, where $r \geq 0$ denotes the risk-free interest rate. Then substituting the expressions of $\Delta$ and $\Delta_1$ in Eq. (\ref{Delta&Delta_1}) into the expectation of $\delta \Pi$ in Eq. (\ref{deltaPi}) yields
\begin{align}
&\bigg[\frac{\partial V}{\partial t}+\frac{S^2}{2}\bigg(\beta^2L^2+\sigma_S^2+2\rho_1\sigma_S\beta L\bigg)\frac{\partial ^2 V}{\partial S^2}+\frac{1}{2}\sigma^2_L\frac{\partial ^2 V}{\partial L^2}+\bigg(\rho_3\sigma_L\beta L+\rho_2\sigma_L\sigma_S\bigg)S\frac{\partial ^2V}{\partial S\partial L}+rS\frac{\partial V}{\partial S}-rV\bigg]\delta t\nonumber\\
 &-\frac{\partial V}{\partial L} \cdot \frac{\bigg[\frac{\partial V_1}{\partial t}+\frac{S^2}{2}\bigg(\beta^2L^2+\sigma_S^2+2\rho_1\beta\sigma_S L\bigg)\frac{\partial ^2 V_1}{\partial S^2}+\frac{1}{2}\sigma^2_L\frac{\partial ^2 V_1}{\partial L^2}+\bigg(\rho_3\sigma_L\beta L+\rho_2\sigma_L\sigma_S\bigg)S\frac{\partial ^2V_1}{\partial S\partial L}+rS\frac{\partial V_1}{\partial S}-rV_1\bigg]}{\frac{\partial V_1}{\partial L}}\delta t\nonumber\\
 &-\mathbb{E}[\kappa S|\nu|]=0.
\end{align}
Utilizing the fact that the value of the known option $V_1$ satisfies the following partial differential equation
\begin{align*}
   & \frac{\partial V_{1}}{\partial t}+\frac{S^2}{2}(\beta^2 L^2+\sigma_S^2+2\rho_{1}\sigma_S\beta L)\frac{\partial^2 V_{1}}{\partial S^2}+\frac{1}{2}\sigma_L^2\frac{\partial^2 V_{1}}{\partial L^2}+(\rho_{3}\sigma_L\beta L+\rho_{2}\sigma_S\sigma_L)S\frac{\partial^2 V_{1}}{\partial S \partial L}+rS\frac{\partial V_{1}}{\partial S}\\
    &-rV_{1}+\alpha(\theta-L)\frac{\partial V_{1}}{\partial L}=0,
\end{align*}
we obtain
\begin{align}
        &\bigg[\frac{\partial V}{\partial t}+\frac{S^2}{2}\bigg(\beta^2 L^2+\sigma_S^2+2\rho_{1}\sigma_S\beta L\bigg)\frac{\partial^2 V}{\partial S^2}+\frac{1}{2}\sigma_L^2\frac{\partial^2 V}{\partial L^2}+\bigg(\rho_{3}\sigma_L\beta L+\rho_{2}\sigma_S\sigma_L\bigg)S\frac{\partial^2 V}{\partial S \partial L}+rS\frac{\partial V}{\partial S}\nonumber\\
        &+\alpha(\theta-L)\frac{\partial V}{\partial L}-rV\bigg]\delta t-\mathbb{E}[\kappa S|\nu|]=0.
        \label{PDE-Init}
\end{align}
To calculate the nonlinear transaction costs term, one needs to find the expression of the number of traded stocks $\nu$ for hedging the portfolio in a time step beforehand. Mathematically, $\nu$ can be presented by the difference of the number of stocks in the portfolio $\Delta$ from time $t$ to $t+\delta t$, where
\begin{align*}
       & \Delta_{t}=\frac{\partial V}{\partial S}(S, L, t)-\Delta_{1}(S, L, t)\frac{\partial V_1}{\partial S}(S, L, t),\\
        &\Delta_{t+\delta t}=\frac{\partial V}{\partial S}(S+\delta S, L+\delta L, t+\delta t)-\Delta_{1}(S+\delta S,L+\delta L, t+\delta t) \frac{\partial V_1}{\partial S}(S+\delta S,L+\delta L, t+\delta t).
\end{align*}
Since the time interval is assumed to be small, Taylor's expansion of $\Delta_{t+\delta t}$ can be adopted to obtain the expression of $\nu$. In addition, with $\delta S=\beta L S\delta W^\gamma_t+\sigma_S S \delta W^S+O(\delta t)$ and $\delta L=\sigma_L \delta W^L+O(\delta t)$, the dominant term of $\nu$ is $O(\delta t)$. Thus, if we only keep the terms of order $O(\sqrt{\delta t})$ and omit all other terms, we can obtain
\begin{align}
 \nu=\Delta_{t+\delta t}-\Delta_{t}=&\beta LS(\frac{\partial^2 V}{\partial S^2}-\Delta_{1}\frac{\partial^2 V_{1}}{\partial S^2}-\frac{\partial \Delta_{1}}{\partial S}\frac{\partial V_{1}}{\partial S})\delta W^{\gamma}\nonumber\\
 &+\sigma_S S(\frac{\partial^2 V}{\partial S^2}-\Delta_{1}\frac{\partial^2 V_{1}}{\partial S^2}-\frac{\partial \Delta_{1}}{\partial S} \frac{\partial V_{1}}{\partial S})\delta W^{S} \\
       &+\sigma_L(\frac{\partial^2 V}{\partial S\partial L}-\Delta_1\frac{\partial^2 V_{1}}{\partial S \partial L}-\frac{\partial \Delta_{1}}{\partial L}\frac{\partial V_{1}}{\partial S})\delta W^{L}\nonumber
\end{align}
It should be noted that the particular option $V_1$ introduced to hedge the risk brought by market liquidity should not impose any effects on the price of the target option $V$. If one carefully observes the PDE governing $V$ (\ref{PDE-Init}), it is clear that $\mathbb{E}[\kappa S|\nu|]$ is the only term that may depend on $V_1$. Thus, if we further assume that the transaction costs term is free from the value of the introduced option $V_1$, then we can ensure that there is no dependence between the value of the target option $V$ and the introduced $V_1$. This means that the trading numbers of the options, $\Delta$ and $\Delta_1$, are respectively dependent on $V$ and $V_1$, which prompts us to assume
\begin{equation*}
\begin{cases}
\displaystyle\Delta_{1}\frac{\partial^2 V_{1}}{\partial S^2}+\frac{\partial \Delta_{1}}{\partial S}\frac{\partial V_{1}}{\partial S}=0,\\
\displaystyle\Delta_1\frac{\partial^2 V_{1}}{\partial S \partial L}+\frac{\partial \Delta_{1}}{\partial L}\frac{\partial V_{1}}{\partial S}=0.
\end{cases}
\end{equation*}
Mathematically, this assumption is equivalent to
\begin{equation}
    \Delta_{1}\frac{\partial V_1}{\partial S}=f(t),
\end{equation}
where $f$ is a function of time. One can easily deduce that $\Delta_1$, as the trading number of another option $V_1$, has nothing to do with the trading number $\Delta$ of the target option as well as its price $V$. Such an assumption makes sense financially since $V_1$ is used to hedge the resulting risk from the introduction of market liquidity, while the primary risk associated with the stock itself has already been hedged with the underlying stock. With this assumption, the number of traded stocks after $\delta t$ can be calculated as follows
\begin{equation}
\label{nu1}
    \nu=\beta LS\frac{\partial^2 V}{\partial S^2}\delta W^{\gamma}+\sigma_S S\frac{\partial^2 V}{\partial S^2}\delta W^{S}+\sigma_L\frac{\partial^2 V}{\partial S\partial L}\delta W^{L}
\end{equation}
Now we are ready to calculate the transaction costs term. Since the three Brownian motions in the above equation are correlated with the correlation structure specified in Equation (\ref{eqcorr}), we can write
\begin{align*}
        \delta W^{\gamma}&=\sqrt{\delta t}Z_{1},\\
        \delta W^{S}&=\rho_{1}\sqrt{\delta t}Z_{1}+\sqrt{1-\rho_{1}^2}\sqrt{\delta t}Z_{2},\\
        \delta W^{L}&=\rho_{3}\sqrt{\delta t}Z_{1}+\frac{\rho_{2}-\rho_{1}\rho_{3}}{\sqrt{1-\rho_{1}^2}}\sqrt{\delta t}Z_{2}+\sqrt{1-\rho_{3}^2-\frac{(\rho_{2}-\rho_{1}\rho_{3})^2}{1-\rho_{1}^2}}\sqrt{\delta t}Z_{3}.
\end{align*}
where $Z_1, Z_2, Z_3 \sim \mathcal{N}(0,1)$ are three independent normal random variables. For the convenience of calculations, we let
\begin{equation*}
\nu=\phi\delta W^{\gamma}+\psi_{1}\delta W^{S}+\psi_{2}\delta W^{L}
\end{equation*}
with
\begin{equation}
\phi=\beta LS\frac{\partial^2 V}{\partial S^2}, \psi_{1}=\sigma_S S\frac{\partial^2 V}{\partial S^2} \ \text{and} \ \psi_{2}=\sigma_L\frac{\partial^2 V}{\partial S \partial L}.
\label{phis}
\end{equation}
Then we substitute the above expressions into $\nu$ in Eq. (\ref{nu1}), which gives
\begin{equation*}
\nu=[(\phi+\rho_{1}\psi_{1}+\rho_{3}\psi_{2})Z_{1}+(\sqrt{1-\rho_{1}^2}\psi_{1}+\frac{\rho_{2}-\rho_{1}\rho_{3}}{\sqrt{1-\rho_{1}^2}}\psi_{2})Z_{2}+\sqrt{1-\rho_{3}^2-\frac{(\rho_{2}-\rho_{1}\rho_{3})^2}{1-\rho_{1}^2}}\psi_{2}Z_{3}]\sqrt{\delta t},
\end{equation*}
and the expected transaction costs during a time interval is given by
\begin{equation}
        E(k_{TC}S|\nu|)=\sqrt{\frac{2\delta t}{\pi}}\kappa S\bigg[\phi^2+\psi_1^2+\psi_2^2+2\rho_{1}\phi\psi_{1}+2\rho_{2}\varphi_{1}\psi_{2}+2\rho_{3}\phi\psi_{2}\bigg]^\frac{1}{2}.
\end{equation}
Substituting the expectation of transaction costs term into Eq. (\ref{PDE-Init}), we can claim that the holding value of an American put option $V^h$ in the holding region ($S\in[S_f, \infty]$) should satisfy $\mathcal{L}^h V^h=0$, where
\begin{align}
\label{holder}
\mathcal{L}^hV^h=&\frac{\partial V^h}{\partial t}+\frac{S^2}{2}(\beta^2 L^2+\sigma_S^2+2\rho_{1}\sigma_S\beta L)\frac{\partial^2 V^h}{\partial S^2}+\frac{1}{2}\sigma_L^2\frac{\partial^2 V^h}{\partial L^2}+(\rho_{3}\sigma_L\beta L+\rho_{2}\sigma_S\sigma_L)S\frac{\partial^2 V^h}{\partial S \partial L}+rS\frac{\partial V^h}{\partial S}\nonumber\\
&+\alpha(\theta-L)\frac{\partial V^h}{\partial L}-rV^h-\sqrt{\frac{2}{\pi\delta t}}\kappa S\bigg[\phi^2+\psi_{1}^2+\psi_{2}^2+2\rho_{1}\phi\psi_{1}+2\rho_{2}\psi_{1}\psi_{2}+2\rho_{3}\phi\psi_{2}\bigg]^\frac{1}{2}.
\end{align}
Here, $S_f$ is the optimal exercise price, and the expressions of $\phi, \psi_1$ and $\psi_2$ are specified in Eq. (\ref{phis}).

Similarly, the writing value of the option $V^w$ in the holding region, which is determined by the holder, should satisfy $\mathcal{L}^w V^w=0$, where
\begin{align}
\label{writer}
\mathcal{L}^wV^w=&\frac{\partial V^w}{\partial t}+\frac{S^2}{2}(\beta^2 L^2+\sigma_S^2+2\rho_{1}\sigma_S\beta L)\frac{\partial^2 V^w}{\partial S^2}+\frac{1}{2}\sigma_L^2\frac{\partial^2 V^w}{\partial L^2}+(\rho_{3}\sigma_L\beta L+\rho_{2}\sigma_S\sigma_L)S\frac{\partial^2 V^w}{\partial S \partial L}\nonumber\\
&+rS\frac{\partial V^w}{\partial S}+\alpha(\theta-L)\frac{\partial V^w}{\partial L}-rV^w+\sqrt{\frac{2}{\pi\delta t}}\kappa S\bigg[\phi^2+\psi_{1}^2+\psi_{2}^2+2\rho_{1}\phi\psi_{1}+2\rho_{2}\psi_{1}\psi_{2}+2\rho_{3}\phi\psi_{2}\bigg]^\frac{1}{2}.
\end{align}
It is obvious that the sign of the transaction costs term in Eq. (\ref{writer}) corresponding to the option writing price is different from that in Equation (\ref{holder}) for the option holding price.

\section{Numerical experiments and examples}
In this subsection, an alternating direction implicit scheme is applied to solve Eqs. (\ref{holder}) and (\ref{writer}) numerically with appropriate boundary conditions. To validate our formulations, the European option prices with zero transaction costs are compared with the closed-form solution in \cite{pasricha2022closed}. Besides, the American option prices with non-zero transaction costs are compared with the results computed by the explicit finite difference method. All of the calculations are carried out for the following parameters unless otherwise mentioned: $S_0=8$, $L_0=0.3$, $K=10$, $r=0.02$, $\beta=0.4$, $\sigma_S=0.3$, $\alpha=2$, $\bar{\theta}=0.6$, $\sigma_L=0.2$, $\rho_1=0.2$, $\rho_2=0.5$, $\rho_3=0.3$, $\lambda=5$, $\zeta=0.5$, $T=1$, $\delta t = \frac{1}{12}$(hedging monthly).
\subsection{Terminal and Boundary conditions}
Since the optimal exercise price is unknown and needs to be determined together with the American option price, the corresponding American option pricing problem can be formed as a LCP defined on $S\in [0, \infty)$. Following \cite{brennan1977valuation}, the American option price for the holder satisfies the variational inequality below 
\begin{equation}
\max (\mathcal{L}^h V^h, \max(K-S, 0)-V^h)=0.
\end{equation}

The terminal condition for an American put option is given by the payoff function: $V(S, L, T)=\max(K-S, 0)$. As the stock price increases to an extremely large value, the put option is worthless, i.e. $\lim_{S\to \infty} V^h(S, L, t)=0$. Also, when the market illiquidity level is already very high, a further increase would only result in a very small change in option prices, which indicates that the first-order derivative of option prices with respect to the illiquidity level should be zero, i.e. $\lim_{L\to \infty} \frac{\partial V^h}{\partial L}(S, L, t)=0$. We also impose the smooth pasting condition as $\frac{\partial V^h}{\partial S}(S_f(t),L, t)=-1$. Moreover, when the illiquidity level approaches zero, the following degenerate boundary is applied from the mathematical point of view with $a\in \{-1, 1\}$:
\begin{equation}\label{sign}
\mathcal{L}^0V=\frac{\partial V}{\partial t}+\frac{\sigma_S^2S^2}{2}\frac{\partial^2 V}{\partial S^2}+\frac{\sigma_L^2}{2}\frac{\partial^2 V}{\partial L^2}+\rho_{2}\sigma_{S}\sigma_{L}S\frac{\partial^2 V}{\partial S \partial L}+rS\frac{\partial V}{\partial S}+\alpha\theta\frac{\partial V}{\partial L}-rV-a*\sqrt{\frac{2}{\pi\delta t}}\kappa S\sqrt{\psi_1^2+\psi_2^2+2\rho_2\psi_1\psi_2},
\end{equation}
where $a=1$ represents the holder, and $a=-1$ stands for the writer. From the mathematical point of view, the sign of the Fichera function $\alpha\theta-\rho_2\sigma_S\sigma_L$  is not guaranteed. Therefore, when $\alpha\theta-\rho_2\sigma_S\sigma_L<0$,  it becomes necessary to prescribe a boundary condition along $L=0$ \cite{fichera1949}. However, while the Fichera function does not specify what particular boundary conditions should be prescribed, we then apply the degenerate boundary condition to the pricing PDEs (\ref{holder}) and (\ref{writer}), as their characteristic forms vanish when $L$ approaches zero.

To sum up, the American option price for the holder  should satisfy
\begin{equation}
\label{holder-bc}
\begin{cases}
\max (\mathcal{L}^h V^h, \max(K-S, 0)-V^h)=0, \\
\frac{\partial V^h}{\partial S}(S_f(t),L, t)=-1,\\
\displaystyle\lim_{S\to \infty} V^h(S, L, t)=0,\\
\mathcal{L}^0 V^h=0 \ \text{for}\  L \to 0,\\
\displaystyle\lim_{L\to \infty} \frac{\partial V^h}{\partial L}(S, L, t)=0,\\
V^h(S, L, T)=\max(K-S, 0).
\end{cases}
\end{equation}

Particularly,  as stated before, there exist slight differences between the pricing of American options for the holder and that for the writer. First, the sign of the transaction costs term differs, as shown in Eq. (\ref{sign}), where $a=1$ represents the holder and $a=-1$ stands for the writer.  Second, the optimal exercise boundary is determined solely by the holder, who possesses the right to exercise the option, whereas the writer must fulfill the obligation. Since this moving boundary has already been determined in Eq. (\ref{holder-bc}) along with the option holding price, the option writing price in both the holding and exercise regions can be computed accordingly. Similarly, the valuation of the American put for the writer can be formulated as the following PDE system
\begin{equation}
\label{writer-bc}
\begin{cases}
	\mathcal{L}^w V^w=0\ \text{for} \ S\in[S_f(t), \infty),\\
   \displaystyle  \lim_{S\to \infty} V^w(S, L, t) = 0,\\
	V^w(S_f(t),L, t)=K-S_f(t),\\
	\mathcal{L}^0 V^w=0 \ \text{for}\  L \to 0,\\
\displaystyle\lim_{L\to \infty} \frac{\partial V^w}{\partial L}(S, L, t)=0,\\
	V^w(S, L, T)=\max(K-S, 0).
\end{cases}
\end{equation}
It should be noted that the PDE systems (\ref{holder-bc}) and (\ref{writer-bc}) are highly nonlinear due to the transaction costs as well as the unknown moving boundary, which do not admit analytical solutions. An efficient and stable numerical scheme is applied to solve such complicated PDE systems numerically, the details of which are presented in the next subsection.

\subsection{The numerical scheme}
Let $V^n_{i, j}= V(S_i, L_j, \tau_n)$ denote the American put option price at time to expiry $\tau_n$ when the stock price is $S_i$ and the liquidity risk is $L_j$. ${S_f}_j^n=S_f(L_j,\tau_n)$ is the optimal exercise price at $\tau_n$ with liquidity risk $L_j$. A uniform grid of $N_S \times N_L$ with $N_T$ time steps is applied for the numerical scheme as listed in Eq. (\ref{def-grid}) of Appendix \ref{AppenA}.

By applying the Douglas-Rachford finite difference scheme \cite{douglas1956numerical}, the numerical solution advancing from time $\tau_n$ to $\tau_{n+1}$ are achieved through a dimensional splitting strategy. The original two-dimensional spatial problem is decomposed into two sequential sub-steps: the first sub-step treats the $S$-direction implicitly; the second sub-step treats the $L$-direction implicitly. This splitting reduces computational complexity by requiring only two inversions of tridiagonal matrix per time step, as opposed to calculating the inverse of a nine-diagonal matrix that would arise in a fully implicit scheme. Specifically, with the initial condition $V^1_{i,j}=\max(K-S_i, 0)$, our pricing PDEs (\ref{holder}) and (\ref{writer}) can be solved backward with the following two sub-steps for $n=1, 2, \cdots, N_T-1$:
\begin{itemize}
\item Sub-step 1: from time $\tau_n$ to $\tau_{n+\frac{1}{2}}$
\begin{equation}
\label{step1}
(I-\eta \Delta \tau A_1)V^{n+\frac{1}{2}}=V^n+(1-\eta)\Delta \tau A_1 V^n+A_0\Delta \tau V^n+A_2\Delta \tau V^n;
\end{equation}
\item Sub-step 2: from time $\tau_{n+\frac{1}{2}}$ to $\tau_{n+1}$
\begin{equation}
\label{step2}
(I-\eta \Delta \tau A_2)V^{n+1}=V^{n+\frac{1}{2}}-\eta \Delta \tau A_2 V^n.
\end{equation}
\end{itemize}
Here, $A_0$ represents the cross derivative term as well as the nonlinear transaction cost term, $A_1$ denotes the spatial derivatives in $S$-direction and $A_2$ stands for the spatial derivatives in $L$-direction. The $rV$ term is split into $A_1$ and $A_2$ equally. $\eta$ is a parameter controlling the type of weighting in the implemented scheme, and it is set to be $\displaystyle\frac{1}{2}$, which produces the Crank-Nicolson scheme, throughout our computations.

It should be pointed out that in the degenerate boundary when $L$ approaches zero, standard finite-difference approximations for the second derivative and cross derivative may require grid points outside the computational domain. To maintain consistency while avoiding out-of-bound indices, these terms are discretized according to Taylor's expansion as follows:
\begin{align}
    \bigg(\frac{\partial ^2 V}{\partial L^2}\bigg)^n_{i, 1}&=\frac{2V^n_{i,0}-5V^n_{i,1}+4V^n_{i,2}-V^n_{i,3}}{(\Delta L)^{2}}+o(\Delta L),\nonumber\\
  \bigg(\frac{\partial^2 V}{\partial S\partial L}\bigg)^n_{i, 1}&=\frac{V^n_{i+1, 2}-V^n_{i+1,1}-V^n_{i-1,2}+V^n_{i-1,1}}{2\Delta S \Delta L}+o(\Delta L\Delta S).
\end{align}
Then, we numerically solve the LCP corresponding to the PDE system (\ref{holder-bc}) using the following ADI scheme that combines central differencing for spatial derivatives with forward differencing for temporal discretization. At each time step, we enforce the early exercise constraint through a projection method, setting the option value to the maximum of either the computed PDE solution or the payoff function.

\begin{breakablealgorithm}
\renewcommand{\algorithmicrequire}{\textbf{Input:}}
\renewcommand{\algorithmicensure}{\textbf{Output:}}
\caption{Numerical solution for the option holding price}
\label{alg:1}
\begin{algorithmic}
 \STATE Given initial condition: ${V^h}^{(1)}=\max(K-S, 0)$.
 \STATE Given initial optimal exercise price: $S_f(L_j, \tau_1)=K$.
\FOR{$n=1, 2, \cdots, N_T-1$}
\STATE Set boundary conditions for ${V^h}^{(n+1)}$.
\STATE Set boundary conditions for ${V^h}^{(n+\frac{1}{2})}$.
\STATE Obtain ${V^h}^{(n+1)}$ by solving Eqs. (\ref{step1}) and (\ref{step2}) as
\STATE $\mathcal{B}{V^h}^{(n+\frac{1}{2})}={\bf{RHS}}^{(n)}+\bf{b}$
\STATE $\mathcal{C}{V^h}^{(n+1)}={\bf{RRHS}}^{(n)}+\bf{c}$
\IF {${V^h}^{(n+1)}\leq (K-S)^+$}
\STATE ${V^h}^{(n+1)}=(K-S)^+$
\ENDIF
\STATE $S_f(L_j,\tau_{n+1})=S\bigg(find\big(|V^h(S_i, L_j, \tau_{n+1})-K+S_i|<10^{-8}, 1, 'last'\big)\bigg)$
\ENDFOR
\end{algorithmic}
\end{breakablealgorithm}
For convenience, the matrices $\mathcal{B}, \mathcal{C}$ and vectors ${\bf{RHS}}^{(n)}, {\bf{RRHS}}^{(n)}, \bf{b}, \bf{c}$ are presented in the Appendix \ref{AppenA}. To efficiently compute the inverses of tridiagonal matrices $\mathcal{B}$ and $\mathcal{C}$, we employ LU decomposition which decomposes the matrix into a lower triangular matrix and an upper triangular matrix to reduce computational complexity. In addition, the optimal exercise boundary $S_f$ is implicitly determined as the largest stock price where the option value equals its intrinsic value. This boundary can be numerically determined through various efficient root-finding methods, such as the bisection method \cite{ehiwario2014, burden2015}, Newton's method \cite{shirzadi2021}, and  iterative methods \cite{dehghan2022, cordero2007}. Then, for a given liquidity risk level, the optimal exercise boundary $S_f(\tau;L)$ divides the $(S, \tau)$-plane into two regions: the holding region $(S\in (S_f(\tau), \infty))$ and the exercise region $(S\in [0, S_f(\tau)])$. Once the optimal exercise boundary is deteremined by the holder, we can accordingly compute the option writing price in each region using the following ADI scheme:

\begin{breakablealgorithm}
\renewcommand{\algorithmicrequire}{\textbf{Input:}}
\renewcommand{\algorithmicensure}{\textbf{Output:}}
\caption{Numerical solution for the option writing price}
\label{alg:1}
\begin{algorithmic}
 \STATE Given initial condition: ${V^w}^{(1)}=\max(K-S, 0)$.
\FOR{$n=1, 2, \cdots, N_T-1$}
\STATE Set boundary conditions for ${V^w}^{(n+1)}$.
\STATE Set boundary conditions for ${V^w}^{(n+\frac{1}{2})}$.
\IF { $S_i \in \bigg[0, S_f(L_j, \tau_n)\bigg]$}
\STATE ${V^w}^{(n+1)}=(K-S)^+$.
\ELSE
\STATE Obtain ${V^w}^{(n+1)}$ by solving Eqs. (\ref{step1}) and (\ref{step2}).
\ENDIF
\ENDFOR
\end{algorithmic}
\end{breakablealgorithm}

\subsection{Validation and order of convergence of our numerical scheme}
Since the pricing of American options we study is a highly nonlinear problem, it is difficult to find its analytical solution. In order to validate our numerical scheme, we first trace back to the European option pricing without transaction costs, whose closed-form solution has been presented in \cite{pasricha2022closed}. From Table 1, we can find that with appropriate space and time steps, the relative difference is less than $0.78\%$, which partially confirms the correctness of our scheme.

\begin{table}[!h]
\caption{European option prices obtained by the closed-form solution in \cite{pasricha2022closed} and our numerical scheme with $\kappa=0$. The specific space and time steps are in the form of $(N_S, N_L, N_T)$.}
\label{validation}
\vskip 5pt
\centering
\begin{tabular}{|c|c|c|c|c|c|c|c|}
\hline
\rowcolor[HTML]{EFEFEF}
\multicolumn{1}{|c|}{{Stock price}}  & \multicolumn{1}{c|} {closed-form }& \multicolumn{2}{c|}{$(50,50,1000)$} & \multicolumn{2}{c|} {$(75,75,1000)$}    & \multicolumn{2}{c|}{$(100, 100, 1000)$}                  \\ \cline{3-8}
\rowcolor[HTML]{EFEFEF}
  {$S_0$}              &solution & Value     & \% difference          & Value          & \% difference    &Value	& \% difference                                                          \\ \hline
8  & 2.4642  & 2.3946 & 2.8274 & 2.4786 & 0.5785 & 2.4672 & 0.1200 \\\hline
9  & 1.8851 & 1.8398 & 2.3995 & 1.9006 & 0.8197 & 1.8927 & 0.4066  \\\hline
10 & 1.4261 & 1.3532 & 5.1099 & 1.4345 & 0.5817 & 1.4333 & 0.5023 \\\hline
11 & 1.0613 & 1.0056 & 5.2550  & 1.0685  & 0.6710 & 1.0697 & 0.7800  \\\hline
12 & 0.7856 & 0.7314  & 6.9006 & 0.7875 & 0.2436 & 0.7877 & 0.2690\\ \hline
\end{tabular}
\end{table}

We now present a numerical verification of the experimental order of convergence (EOC), which is defined as
\begin{equation*}
	\text{EOC}_{i+2}=\frac{\text{ln Difference}_{i+2}-\text{ln Difference}_{i+1}}{\text{ln} N_{\tau,\,i+1}-\text{ln} N_{\tau,\,i+2}}.
\end{equation*}
Table 2-4 illustrates that our numerical scheme is approximately 2$^{\text{nd}}$ order convergent in time and 4$^{\text{th}}$order convergent in space. To balance the convergence order and computational time, we choose $N_S=N_L=100$ and $N_T=1000$ for the ADI scheme, whereas $N_S=N_L=200$, $N_T=750000$ for the fully explicit scheme for the following calculations.

\begin{table}[H]
\caption{EOC in $\tau$-direction for $S_0=8$, $N_S=100$, $N_L=80$.}
\vskip5pt
\centering
\begin{tabular}{|c|c|c|c|c|c|c|}
\hline
\rowcolor[HTML]{EFEFEF}
\multicolumn{1}{|c|}{No. of steps} & \multicolumn{3}{c|}{Holder price} & \multicolumn{3}{c|}{Writer price}                      \\ \cline{2-7}
\rowcolor[HTML]{EFEFEF}
in $\tau$-direction  & Value     & Difference   & EOC    & Value     & Difference   & EOC \\ \hline
2000 & 2.447803 & -        & -        & 2.574484 & -        & -         \\
3000 & 2.447810  & 7.38E-06 & -        & 2.574476 & 8.11E-06 & -          \\
4000 & 2.447814 & 3.68E-06 & 2.4229  & 2.574471 & 5.17E-06 & 1.5674   \\
5000 & 2.447816 & 2.21E-06 & 2.2897 & 2.574467 & 3.21E-06 & 2.1326   \\
 \hline
\end{tabular}
\label{tau_convergence}
\end{table}


\begin{table}[h!]
\caption{EOC in S-direction for $S_0=8$, $N_T=2000$, $N_L=50$.}
\vskip 5pt
\centering
\begin{tabular}{|c|c|c|c|c|c|c|}
\hline
\rowcolor[HTML]{EFEFEF}
\multicolumn{1}{|c|}{No. of steps} & \multicolumn{3}{c|}{Holder price} & \multicolumn{3}{c|}{Writer price}                      \\ \cline{2-7}
\rowcolor[HTML]{EFEFEF}
 in S-direction  & Value     & Difference   & EOC    & Value     & Difference   & EOC \\ \hline
 20  & 2.4386 & -        & -        & 2.5617 & -        & -         \\
60  & 2.3988 & 0.0399 & -        & 2.5180    & 0.0437 & -          \\
100 & 2.3962 & 0.0025 & 5.4069 & 2.5146 & 0.0034 & 5.0098  \\
140 & 2.3957 & 0.0006 & 4.3683 & 2.5138 & 0.0008 & 4.2911   \\          \hline
\end{tabular}
\label{S_EOC}
\end{table}

\begin{table}[h!]
\caption{EOC in L-direction for $S_0=8$, $N_T=3000$, $N_S=60$.}
\vskip 5pt
\centering
\begin{tabular}{|c|c|c|c|c|c|c|}
\hline
\rowcolor[HTML]{EFEFEF}
\multicolumn{1}{|c|}{No. of steps} & \multicolumn{3}{c|}{Holder price} & \multicolumn{3}{c|}{Writer price}                      \\ \cline{2-7}
\rowcolor[HTML]{EFEFEF}
 in L-direction  & Value     & Difference   & EOC    & Value     & Difference   & EOC \\ \hline
20 & 2.0867 & -  & -        & 2.1063 &  -        &  -          \\
 35 &  2.5652 &  0.4786 &  -        &  2.7304 &  0.6241  &  -          \\
 50 &  2.3988&  0.1664 &  2.9614 &  2.5180 &  0.2124 &  3.0219    \\
 65 &  2.4486  &  0.0498 &  4.5959 &  2.5772 &  0.0592 &  4.8678   \\ \hline
\end{tabular}
\label{L_EOC}
\end{table}

Table 5 displays that the relative computational error between the results from the ADI method and those from the fully explicit method is still less than $0.61\%$ when valuing American options with transaction costs, which demonstrates that our scheme is correct. Moreover, compared to the fully explicit scheme, the ADI scheme shows less computation time\footnote{Note that all of our calculations are carried out using Matlab R2017a under Windows 10, and the CPU time are about 500 and 806 seconds for the ADI and fully explicit method, respectively.} and higher experimental order of convergence, which indicates that the ADI scheme is more superior in terms of computational efficiency. As is well known, the ADI scheme is unconditionally stable for linear PDE systems. When applied to nonlinear PDEs, however, stability imposes certain restrictions on the time step. Due to the high nonlinearity, it is challenging to explicitly define these restrictions. To address this, we vary the number of steps in the $\tau$ direction from 1000 to 5000 and observe that the option price values converge, which numerically demonstrates the stability of our ADI scheme.

\begin{table}[H]
\caption{American option prices computed by the fully explicit scheme and ADI scheme with $\kappa=0.8\%$.}
\vskip 5pt
\centering
\begin{tabular}{|c|c|c|c|c|c|c|}
\hline
\rowcolor[HTML]{EFEFEF}
\multicolumn{1}{|c|}{Stock price} & \multicolumn{3}{c|}{Holder price} & \multicolumn{3}{c|}{Writer price}                      \\ \cline{2-7}
\rowcolor[HTML]{EFEFEF}
 $S_0$  & Explicit     & ADI   & RD(\%)    &  Explicit     & ADI   & RD(\%) \\ \hline
8     & 2.4458   & 2.4469    & 0.0447   & 2.5725   & 2.5735   & 0.0399    \\
9     & 1.8472    & 1.8482   & 0.0560   & 2.0023   & 2.0038   & 0.0711     \\
10    & 1.3682    & 1.3750    & 0.5004   & 1.5386   & 1.5452   & 0.4272     \\
11    & 0.9994    & 1.0056    & 0.6084   & 1.1725    & 1.1783   & 0.4942    \\
12    & 0.7196   & 0.7236    & 0.5571   & 0.8856   & 0.8895   & 0.4351     \\  \hline
\end{tabular}
\label{S_convergence}
\end{table}

\subsection{Optimal exercise boundary}
As a key feature of American options, we analyze how exogenous and endogenous transaction costs affect the optimal exercise boundary. In particular, the influence of significant factors such as proportional transaction costs rate $\kappa$, the mean-reversion speed $\alpha$ and level $\bar{\theta}$ of liquidity risk, and the price sensitivity $\beta$ to market liquidity is investigated through numerical experiments.
\begin{figure}[H]
	\centering  
	\subfigbottomskip=2pt 
	\subfigcapskip=-5pt 
	\subfigure[Different $\kappa$]{
		\includegraphics[width=0.48\linewidth]{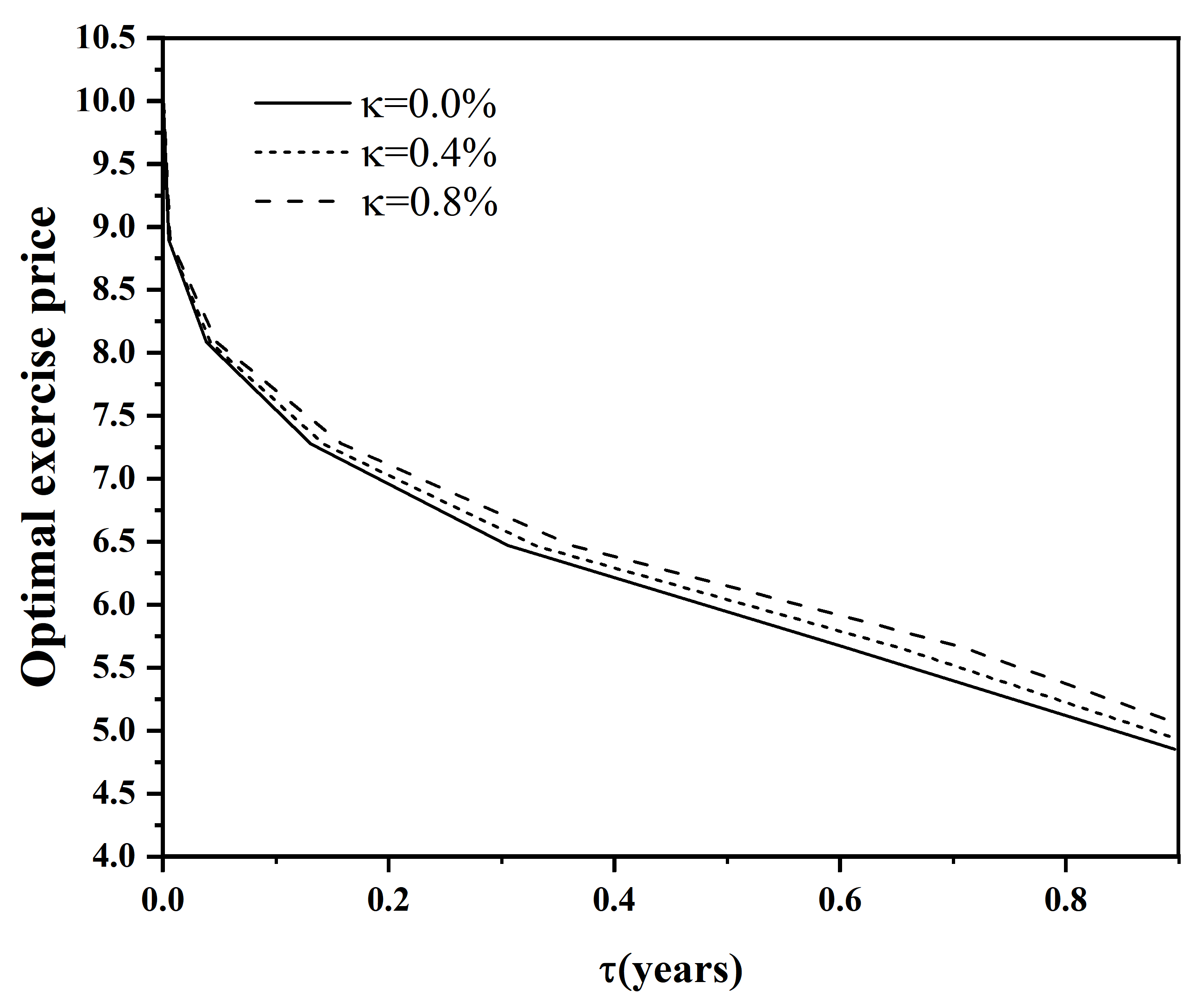}}
	\subfigure[Different $\alpha$]{
		\includegraphics[width=0.48\linewidth]{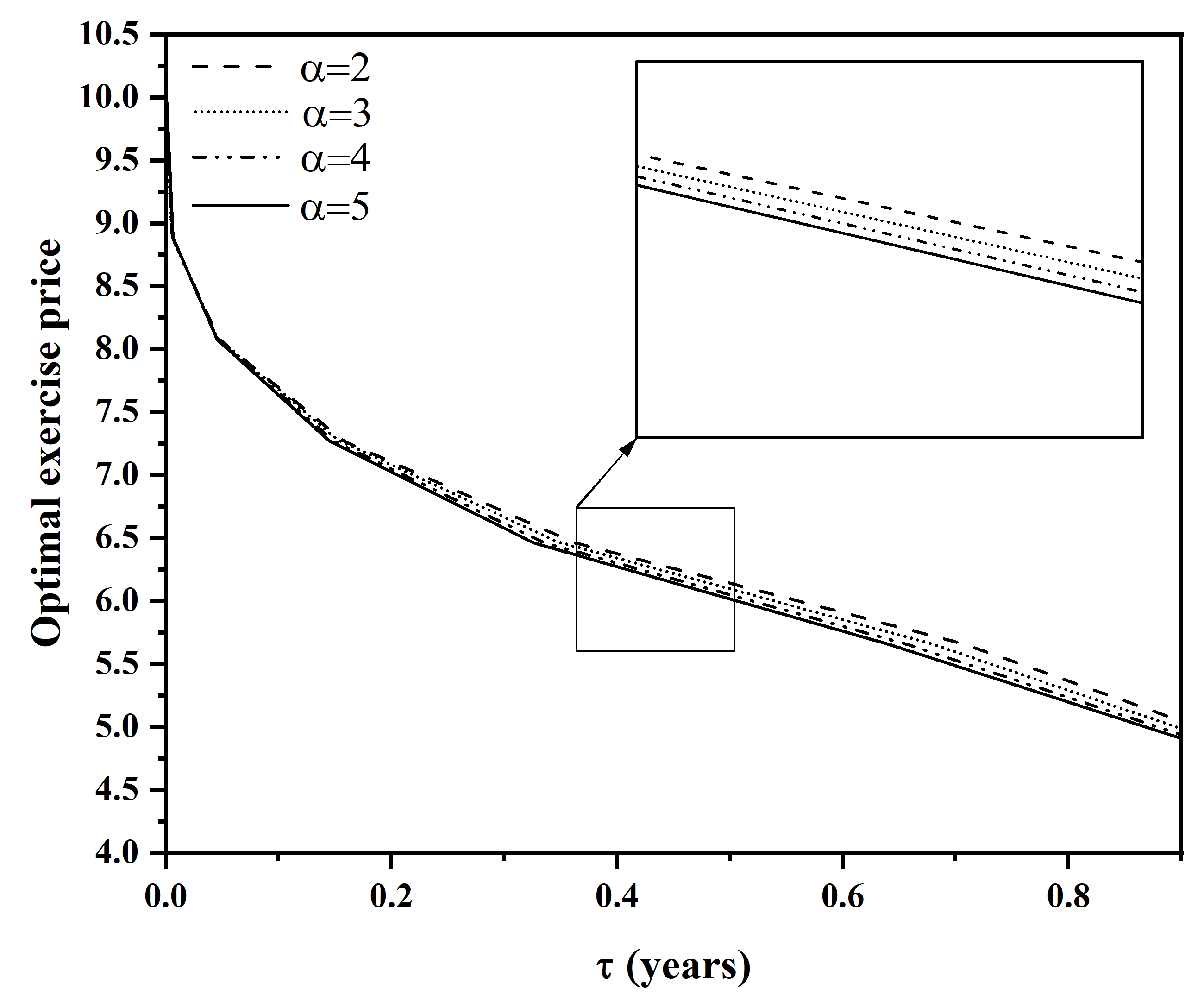}}
	  \\
	\subfigure[Different $\beta$]{
		\includegraphics[width=0.48\linewidth]{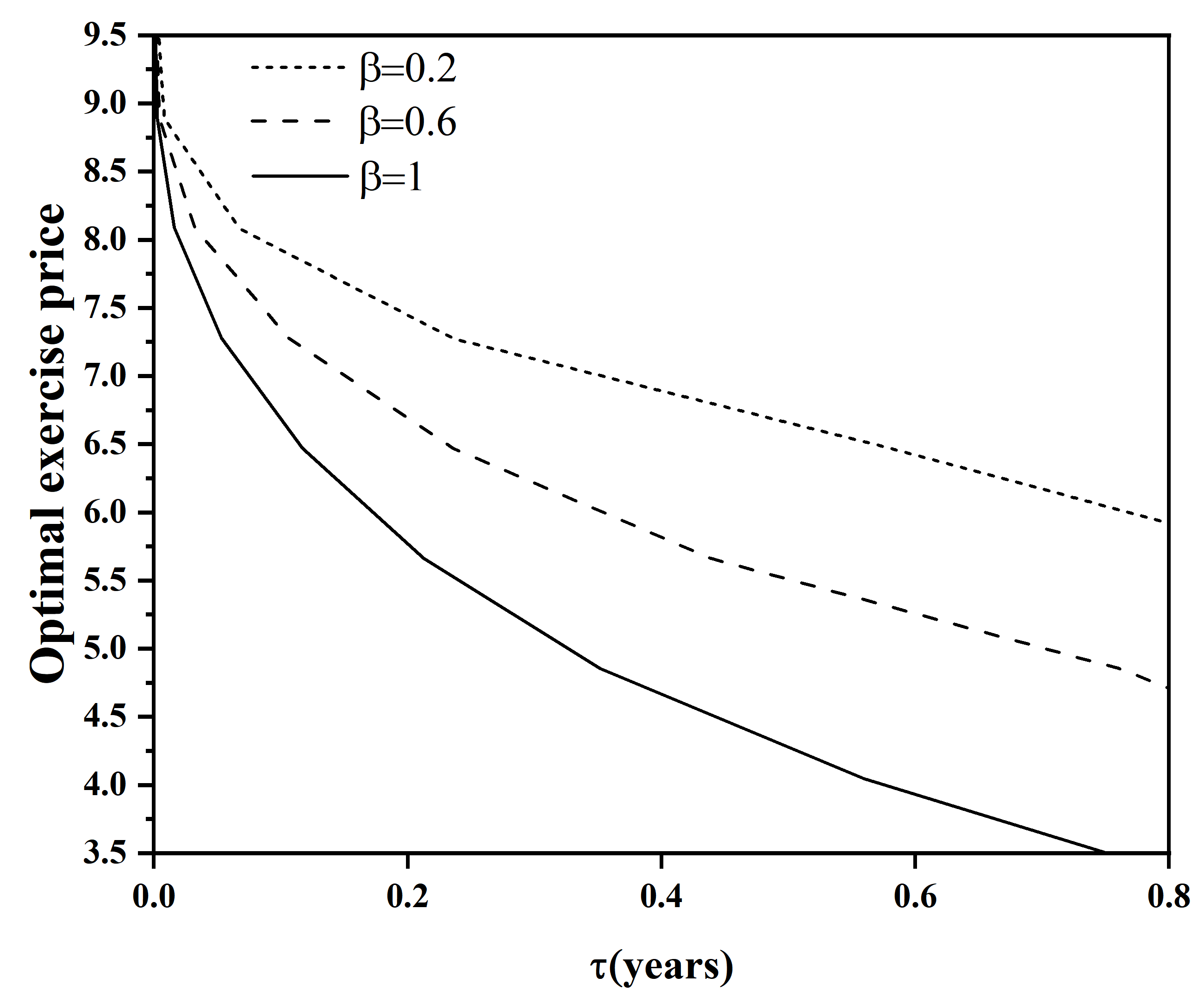}}
	\subfigure[Different $\overline\theta$]{
		\includegraphics[width=0.48\linewidth]{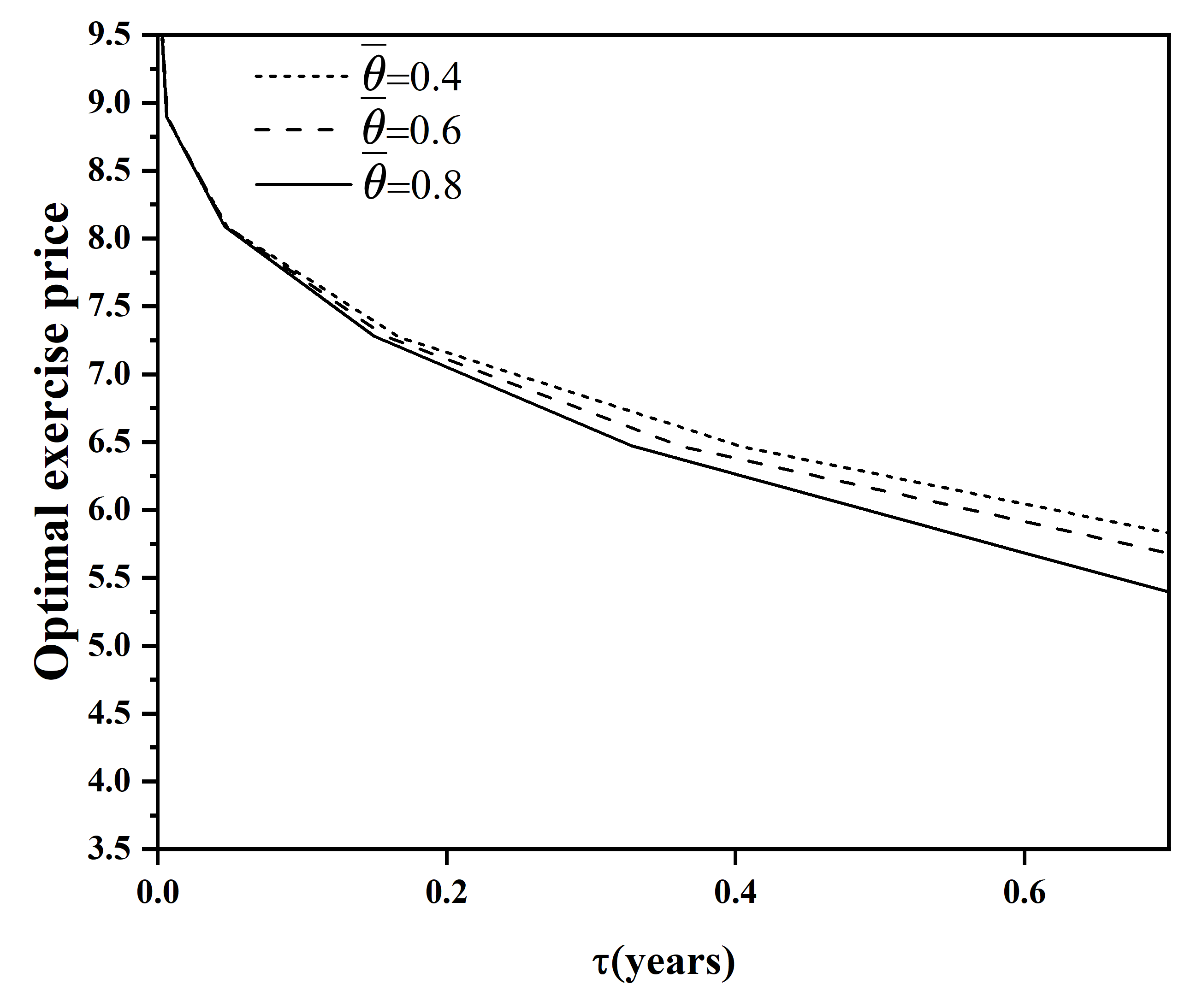}}
\caption{Option exercise price with different parameters.}
\end{figure}

In Figure 1(a), the optimal exercise price is shown to be an increasing function of the proportional transaction costs rate. This is reasonable since when $\kappa$ increases, it would cost the option holder more in the dynamic hedging process, and the holder naturally wants to exercise early to reduce further costs. One can observe from Figure 1(b) that the optimal exercise price decreases when there is an increase in the mean-reversion speed of the market illiquidity. One can understand this from the aspect that under the current parameter settings, a larger mean reversion speed implies a larger average market illiquidity level, which means that the actual stock price is lower than its market price, and thus the holder would like to exercise the option when the stock price reaches a lower level. A similar phenomenon is displayed in Figure 1(d). One can further use this theory to explain what has been shown in Figure 1(c), where it is clear that a larger $\beta$ indicates that the option would be exercised later. This is because that an increase in $\beta$ means that market illiquidity level has more impact on stock prices, which prompts the holder to hold the option for a longer time.

\subsection{American option price}
In this section, we analyze the effect of exogenous and endogenous transaction costs on American option prices.
\begin{table}[h!]
\caption{American option prices for $\kappa_{TC}=0$,  $\kappa_{TC}=0.4\%$ and $\kappa_{TC}=0.8\%$}
\vskip 5pt
\centering
\begin{tabular}{|c|c|c|c|c|c|}
\hline
\rowcolor[HTML]{EFEFEF}
\multicolumn{1}{|c|}{{Stock price}}  & \multicolumn{1}{|c|} {\phantom{No TC}}  & \multicolumn{2}{c|}{Holder} & \multicolumn{2}{c|} {Writer}                      \\ \cline{3-6}
\rowcolor[HTML]{EFEFEF}
{$S$}              &  {$\kappa_{TC}=0$}  &  $\kappa_{TC}=0.4\%$         & $\kappa_{TC}=0.8\%$   &   $\kappa_{TC}=0.4\%$   & $\kappa_{TC}=0.8\%$                                                         \\ \hline
8  & 2.5009& 2.4742 & 2.4469 & 2.5381 & 2.5735   \\
9  & 1.9143 & 1.8818 & 1.8482  & 1.9599  & 2.0037   \\
10 & 1.4473 & 1.4118 & 1.3751 & 1.4971 & 1.5451  \\
11 & 1.0787 & 1.0427 & 1.0058 & 1.1292 & 1.1783   \\
12 & 0.7935 & 0.7590 & 0.7236 & 0.8420 & 0.8895 \\  \hline
\end{tabular}
\label{results_with_TC}
\end{table}

Table 6 illustrates that when there are proportional transaction fees, the option price of the holder is lower than that of the writer, and such an interval is enlarged when there is an increase in the transaction costs rate. This is because both the holder and the writer would like to count in the effects of transaction cost, thus respectively asking for lower and higher option prices. Then we analyze the impact of key liquidity risk parameters $\alpha, \beta, \bar{\theta}$ on the option prices.

\begin{figure}[H]
	\centering
	\subfigbottomskip=2pt
	\subfigcapskip=-5pt
	\subfigure[Different $\alpha$]{
		\includegraphics[width=0.48\linewidth]{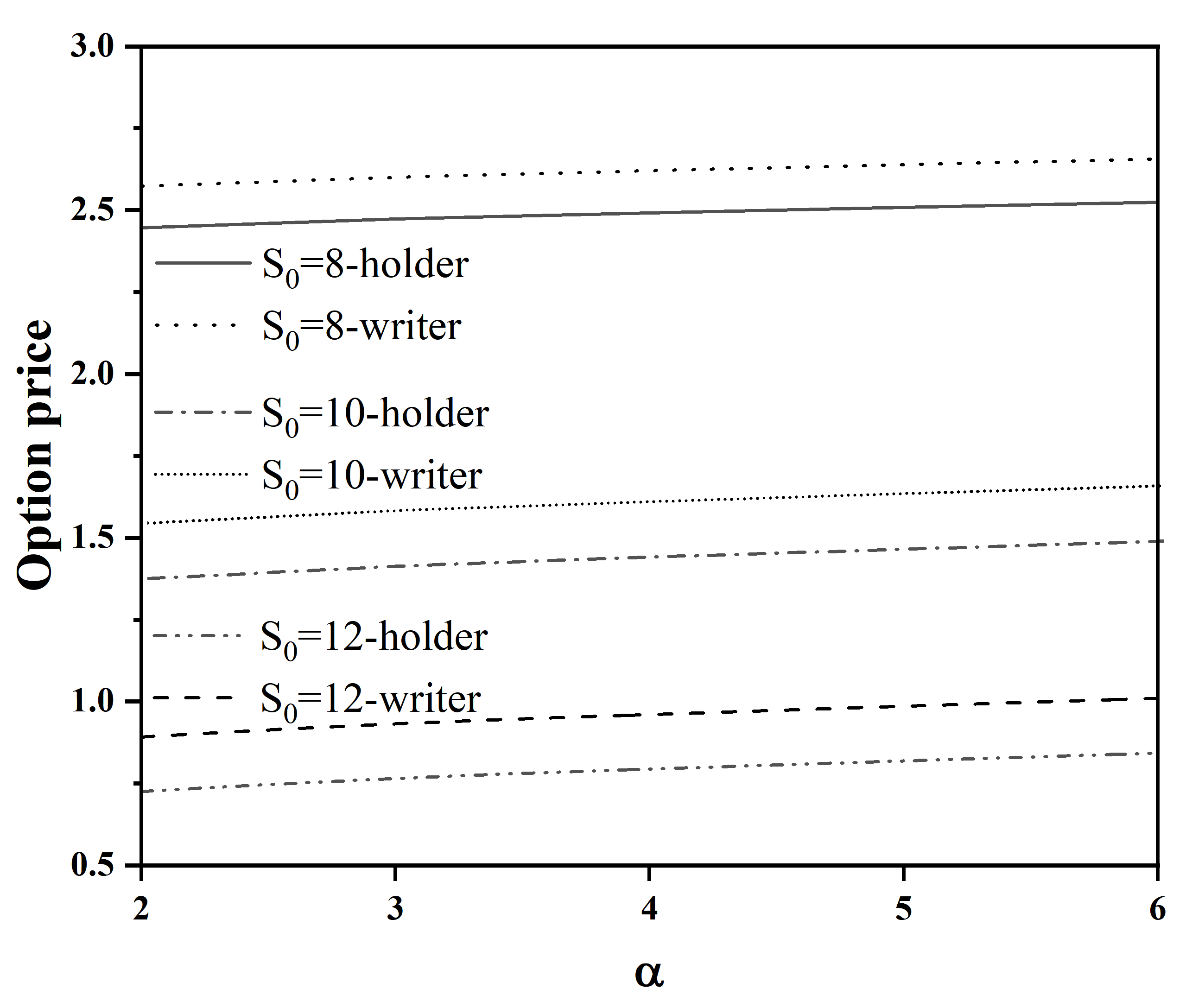}}
	\subfigure[Different $\beta$]{
		\includegraphics[width=0.48\linewidth]{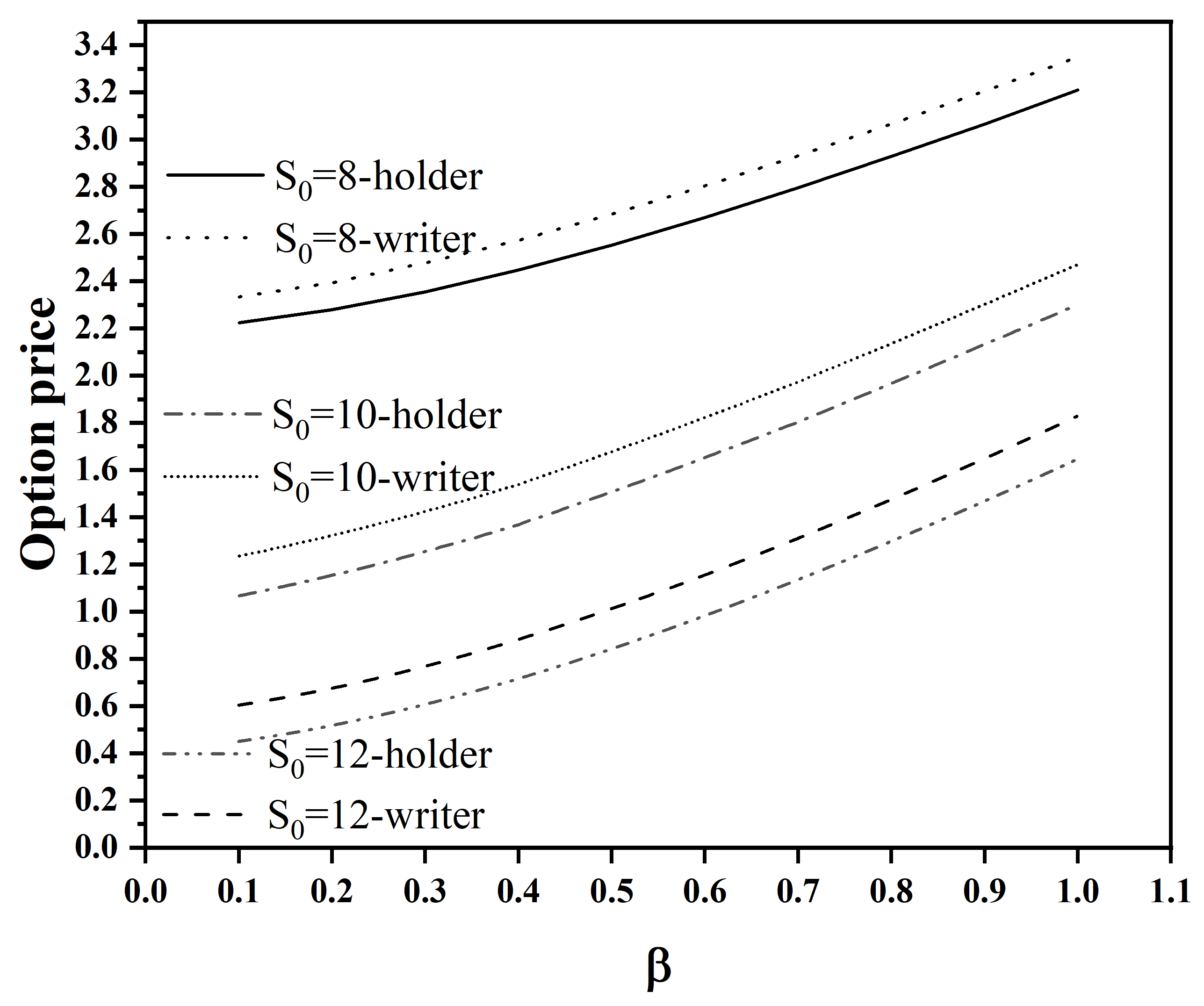}}
	  \\
	\subfigure[Different $\overline\theta$]{
		\includegraphics[width=0.48\linewidth]{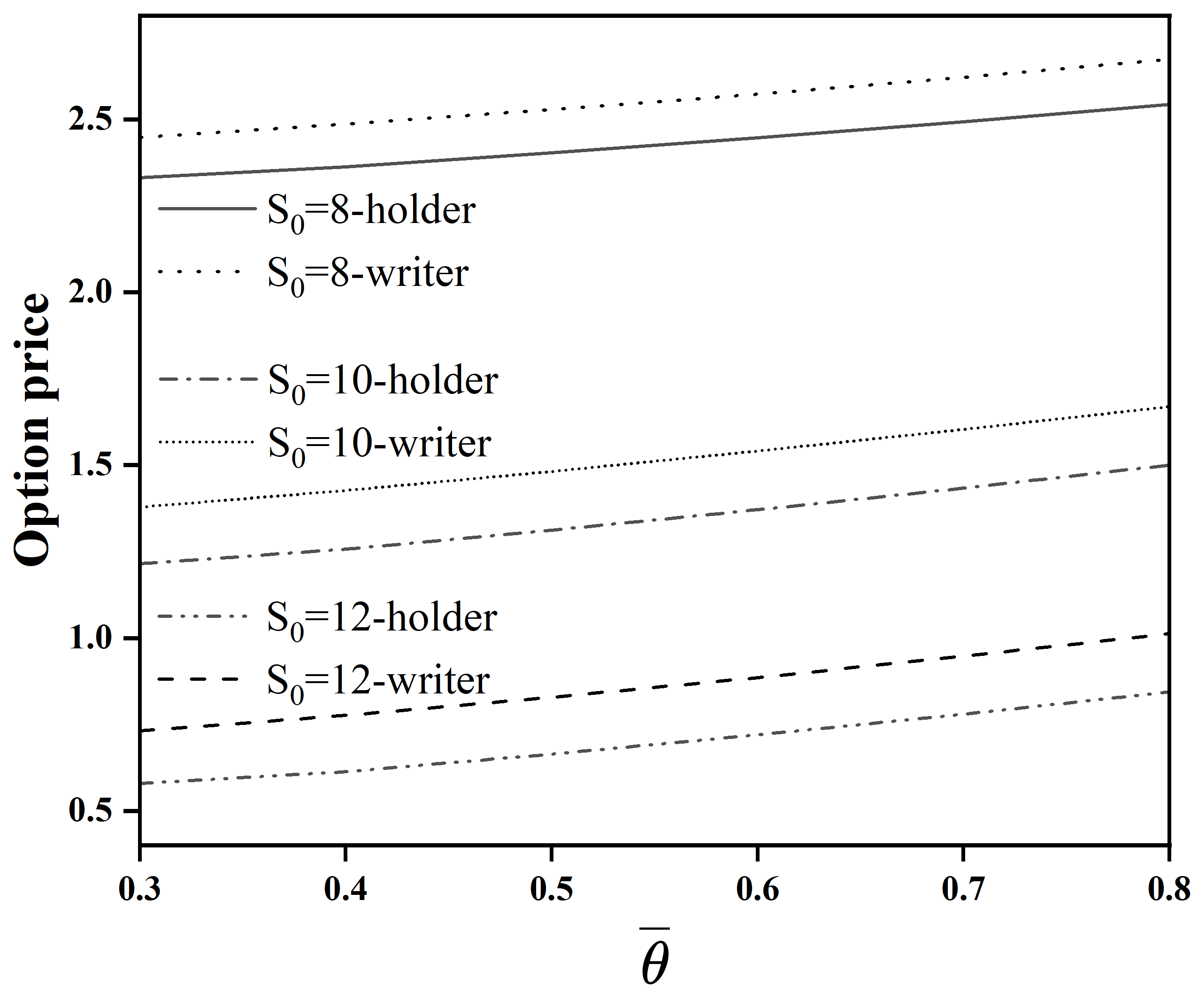}}
\caption{American option prices with different parameters.}
\end{figure}

One can then observe from Figure 2(a) that American put option prices are an increasing function of the mean-reversion speed of the market illiquidity. The main explanation for this is that as discussed above, a larger $\alpha$ means a larger average market illiquidity level, which can actually indicate that the stock price is devalued to a larger extent, thus giving rise to option prices. This also account for the phenomenon shown in Figures 2(b) and 2(c), since larger $\beta$ or larger $\overline\theta$ naturally means that the liquidity level of the stock is worse.

\section{An empirical study}
In this section, we would conduct an empirical study to demonstrate the performance of our proposed model. We choose the Leland model \cite{leland1985option}, which excludes liquidity risk in our pricing PDEs (\ref{holder}) and (\ref{writer}), as the  benchmark model in our empirical study. First, we specify how to select empirical data. Then, parameter estimation is conducted separately for our model and the benchmark model using the MLE method. Finally, the pricing performance is assessed by comparing its root mean square error (RMSE), which measures the effectiveness and importance of incorporating liquidity dynamics into the pricing framework.
\subsection{Data selection}
In our study, we utilize two datasets comprising soybean meal futures and options from the Wind database. These datasets serve as essential inputs for model parameter estimation and empirical analysis. The futures dataset contains daily closing prices of soybean meal futures contracts traded on the Dalian Commodity Exchange, spanning the period from January 2022 to January 2024. The options dataset includes American put options on soybean meal futures, also traded on the Dalian Commodity Exchange. To ensure data quality and liquidity, we apply a filter excluding option contracts with daily trading volumes below 1,200 contracts. This selection criterion guarantees our analysis focuses on actively traded options, thereby enhancing the reliability of our results.

\subsection{Estimation procedure}
In this subsection, we estimate the parameters for both the benchmark model and our proposed model with different MLE methods due to the different dynamics.  Before we present the parameter estimation method, it should be noted that futures markets exhibit short-term volatility that can affect parameter estimation when using fixed window. In order to mitigate the effects of short-term futures market volatility and enhance estimation accuracy, we employ a moving window method, as part of our methodology. In specific, each of our five estimation windows contains 762 trading days, with consecutive windows shifting forward by five trading days. The first window spans from the first trading day of 2022 to the February of 2025. Subsequent windows shift forward accordingly, with the fifth window extending into March 2025. We apply MLE \cite{hamilton2020time} to calibrate our model parameters using empirical data.

\subsubsection{Calibration of the benchmark model}
The benchmark model specifies the underlying asset price process $S_t$ as a geometric Brownian motion with constant drift $\mu$ and volatility $\sigma$:
\begin{equation*}
    d S_t = \mu S_t d t + \sigma S_t d W_t^P,
\end{equation*}
with $W_t^P$ is a standard Brownian motion under the physical measure $P$.

To estimate the parameters $\mu$ and $\sigma$, we employ the standard MLE method. Given discrete observations of underlying asset prices $\{S_t\}$ at time intervals $\Delta t$, the log-likelihood function is constructed as:
\begin{equation}
    \log L(\mu, \sigma) = -\frac{1}{2} \sum_{t} \left[ \log(2\pi \sigma^2 \Delta t) + \frac{(\log S_t - \log S_{t-1} - \mu \Delta t)^2}{\sigma^2 \Delta t} \right].
\end{equation}
The parameter estimates $\hat{\mu}$ and $\hat{\sigma}$ are obtained by maximizing the log-likelihood function.

\subsubsection{Calibration for our model}
Since our model incorporates the unobservable liquidity risk process $L_t$, it is naturally cast within a state-space framework. Given the nonlinear nature of our model's dynamics, the standard Kalman filter is insufficient. We therefore employ the extended Kalman filter (EKF), which handles nonlinearities by performing sequential linearizations at each time step. Using the EKF to compute the likelihood function, which is then maximized to estimate parameters, is a well-established methodology for this class of problems \cite{durbin2012time}.

This EKF-MLE approach has been successfully applied across a wide range of financial applications involving latent variables, such as in affine term structure models \cite{duffee2002term} and for pricing commodity derivatives with unspanned factors \cite{trolle2009unspanned}. Following this established literature, our methodology allows us to recursively estimate the latent liquidity state while simultaneously optimizing the model parameters to fit the observed price dynamics.
Within this estimation framework, the state-space representation of our model is given by:
\textbf{State vector:}
\begin{equation*}
X_t = \begin{bmatrix} S_t,  L_t \end{bmatrix}^{\top},
\end{equation*}
where $S_t$ is the underlying asset price and $L_t$ is the latent liquidity process.

\textbf{Transition equation:}
\begin{equation*}
\begin{cases}
        S_{t+1} = S_t + \mu S_t \Delta t + \beta L_t S_t \Delta W^\gamma_t + \sigma_S S_t \Delta W^S_t,\\
        L_{t+1} = L_t + \alpha(\theta - L_t)\Delta t + \sigma_L \Delta W^L_t,
\end{cases}
\end{equation*}
which describes the discrete-time dynamics of the state variables and can be expressed in the general form:
\begin{equation*}
X_{t+1} = f(X_t) + w_{t+1}, \quad w_{t+1} \sim \text{i.i.d}, \quad E[w_{t+1}] = 0, \quad \text{Var}[w_{t+1}] = Q_t,
\end{equation*}
where $f$ represents the nonlinear transition function derived from our continuous-time dynamics, and $Q_t$ is the covariance matrix of the system noise.

\textbf{Measurement equation:}
\begin{equation*}
    y_t = h(X_t) + \epsilon_t,
\end{equation*}
where $y_t$ represents the observed market price data, and $\epsilon_t \sim \mathcal{N}(0, R)$ is measurement noise. 

The extended Kalman filter (EKF) iteratively updates the state estimates by linearizing the nonlinear transition function at each time step and computes the log-likelihood function, which is then maximized to obtain optimal parameter estimates. In particular, we write
\begin{equation}
    \log L(\Theta) = -\frac{1}{2} \sum_t \left[ \log(2\pi N_t)+\log(|V_t|) + e_t' V_t^{-1} e_t \right],
\end{equation}
where $N_t$ is the dimension of $e_t$, $V_t$ is the covariance of the prediction error, and $e_t$ is the innovation term. The parameter vector $\Theta = (\mu,\alpha,\beta,\bar{\theta} ,\sigma_S, \sigma_L, \rho_1, \rho_2,\rho_3,\lambda,\zeta)$ is estimated by maximizing the log-likelihood function using numerical optimization techniques. 

For our model, we set $\kappa_{TC}=0.005\%$, as it is officially specified by the Dalian Commodity Exchange. While this rate appears negligible at first glance, it significantly impacts pricing dynamics due to the distinctive nature of soybean meal futures trading. Unlike purely speculative markets, participants in soybean meal futures are predominantly commercial entities with genuine commodity demands, resulting in exceptionally large transaction volumes. Consequently, even this minimal percentage translates to substantial absolute costs that influence trading decisions and price formation processes. 

By comparing the estimation results from both models, we systematically assess the impact of liquidity on pricing dynamics and quantify the improvements achieved by incorporating latent liquidity into the stochastic process, providing a more realistic representation of the market's microstructure.

\subsection{Empirical results}

After applying the MLE methods mentioned above, we obtain two set of parameter estimates as shown in Table 7, where t-statistics are in parentheses.
\begin{table}[H]
\centering
\caption{Estimated parameters}
\small
\setlength{\tabcolsep}{3pt}
\renewcommand{\arraystretch}{0.9}
\begin{tabular}{c|ccccccccccc|c}
\toprule
\textbf{Model} & $\mu$ & $\alpha$ & $\beta$ & $\bar{\theta}$ & $\sigma_S$ & $\sigma_L$ & $\rho_1$ & $\rho_2$ & $\rho_3$ & $\lambda$ & $\zeta$ & $- \log L(\mu, \sigma)$ \\
\midrule
Our & $-0.0028$ & $1.9921$ & $0.8161$ & $0.1700$ & $0.1238$ & $0.1237$ & $0.2097$ & $0.5061$ & $0.3086$ & $5.0000$ & $0.5000$ & $-4134.995$ \\
Model & $(-0.03)$ & $(12.75)$ & $(5.26)$ & $(1.09)$ & $(2.5)$ & $(0.92)$ & $(1.41)$ & $(3.24)$ & $(1.98)$ & $(32.01)$ & $(3.2)$ & \\
Benchmark & $-0.0022$ & \multirow{2}{*}{/} & \multirow{2}{*}{/} & \multirow{2}{*}{/} & $0.2527$ & \multirow{2}{*}{/} & \multirow{2}{*}{/} & \multirow{2}{*}{/} & \multirow{2}{*}{/} & \multirow{2}{*}{/} & \multirow{2}{*}{/} & $2070.9849$ \\
Model & $(-0.01)$ & & & & $(39.01)$ & & & & & & & \\
\bottomrule
\end{tabular}
\end{table}

We adopt these estimated parameters shown in Table 7 to predict the theoretical prices for American put option by using our PDE method. These predicted results are then compared against the actual call option price to examine the error in the two models, referred to as out-of-sample error. We adopt the following measure to demonstrate the  performance of the models:
\begin{equation}
    \text{RMSE} = \sqrt{\frac{1}{n}\sum_{i=1}^{n}(P_{\text{theoretical},i} - P_{\text{actual},i})^2}
\end{equation}

\begin{table}[h!]
\caption{Comparison of Out-of Samples Absolute Pricing Error}
\vskip 5pt
\centering
\begin{tabular}{|c|c|c|c|c|}
\hline
\rowcolor[HTML]{EFEFEF}
\multicolumn{1}{|c|}{\textbf{Moneyness}} &\textbf{All} & \textbf{Out-of-money} & \textbf{In-the-money} & \textbf{At-the-money}  \\ \hline
Our model & 4.820383508 & 5.35617311 & 4.261255055 & 4.528801985  \\
Benchmark & 16.59684615 & 15.37334875 & 10.78263422 & 18.62326873  \\ \hline
\end{tabular}
\end{table}

The pricing errors between our model and the benchmark model are shown  across all moneyness categories in Table 8. From this table, we find that our new model outperforms the benchmark model for out-of-money, at-the-money, and in-the-money options. For all options, our model achieves an RMSE of 4.82, which is significantly lower than the benchmark's 16.60. In the out-of-the-money category (\(S/K > 1.03\)), our model's RMSE is 5.36, compared to the benchmark's 15.37, indicating better performance in pricing options that are less likely to be exercised. For in-the-money options (\(S/K < 0.97\)), our model achieves an RMSE of 4.26, while the benchmark's RMSE is 10.78, highlighting our model's effectiveness in pricing options with positive intrinsic values. In the at-the-money category (\(0.97 \leq S/K \leq 1.03\)), our model's RMSE is 4.53, which is significantly lower than the benchmark's 18.62, underscoring the model's robustness in pricing options where the underlying asset price is close to the strike price. Overall, the results clearly indicate that incorporating liquidity risks into the pricing model significantly enhances its accuracy, particularly for at-the-money options, where precision is crucial.

\section{Conclusion}
In this paper, the pricing problem of American options is studied when both exogenous and endogenous transaction costs are taken into consideration. While endogenous transaction costs here are referred to as liquidity risks, which are modeled with an Ornstein-Uhlenbeck process, exogenous transaction costs are associated with the incurred fees in each trading and are assumed to be proportional to the transaction value. The intrinsic connection between the two types of transaction costs is also captured. Due to the existence of the moving boundary together with transaction costs, the resulting pricing PDE systems are highly nonlinear with mixed derivative terms, and are solved with an efficient numerical algorithm. We find that the option holding price is always lower than the option writing price, and such a fair price range is enlarged with the increase in the transaction costs rate. We also discover that the option holder naturally wants to hold the option for more time to possibly get more profits when market illiquidity levels rise, in which case the corresponding American option prices are higher. Our empirical analysis demonstrates that incorporating liquidity risk into the pricing framework significantly improves model accuracy, as measured by reduced pricing errors across all moneyness categories. This improvement is notable for at-the-money options, where pricing precision is crucial. In addition, our modeling framework can be readily adapted to pricing various option types while also being extendable to incorporate additional market factors such as stochastic volatility, stochastic interest rates and jump diffusions.

\section*{Acknowledgments}
Xin-Jiang He acknowledges the financial supports from the National Natural Science Foundation of China (No. 12101554) and the Fundamental Research Funds for Zhejiang Provincial Universities (No. GB202103001). Guiyuan Ma acknowledges the financial supports from the National Natural Science Foundation of China (72101199) and the System Behavior and Management Laboratory of Xi'an Jiaotong University, Philosophy and Social Sciences Laboratory of the Ministry of Education in China. The authors would also like to gratefully acknowledge the anonymous referees' constructive comments and suggestions, which greatly help to improve the quality of the manuscript.

\section*{Data availability statement}
All the codes can be found on GitHub (\url{https://github.com/haixianw/Finite_Difference_Method.git}).

\begin{appendices}
\section{Grid and coefficients for numerical scheme in Section 3}\label{AppenA}
The uniform grid for our numerical scheme is defined as follows:
\begin{align}
\label{def-grid}
&\tau_n=(n-1)\Delta \tau, n=1, 2, \cdots, N_T, \Delta \tau=\frac{T}{N_T-1},\nonumber\\
&S_i=(i-1)\Delta S, i=1, 2, \cdots, N_S, \Delta S=\frac{S_{max}-S_{min}}{N_S-1},\\
&L_j=(j-1)\Delta L, j=1, 2, \cdots, N_L, \Delta L=\frac{L_{max}-L_{min}}{N_L-1},\nonumber
\end{align}
where $S_{min}=L_{min}=0$, $S_{max}=8K, L_{max}=5.$

To compute Eq. (\ref{step1}), the tridiagonal matrix $\mathcal{B}$ with size $[(N_S-2)\cdot (N_L-2)] \times [(N_S-2)\cdot (N_L-2)]$, and vectors $\bf{RHS}, \bf{b}$ are given by
\begin{equation*}
 \mathcal{B}=\begin{pmatrix}
  D_{2,2}&-E_{2,2} &  & &\\
 - F_{3,2}& D_{3,2}  &  && \\
  &&  \ddots& \ddots & \\
 &&  \ddots& \ddots &-E_{N_{S-2},N_{L-1}} \\
 &&  &-F_{N_{S-1},N_{L-1}} & D_{N_{S-1},N_{L-1}}
\end{pmatrix},
\end{equation*}
\begin{equation*}
 {\bf{RHS}}^{(n)}=(R_{2,2}^n,R_{3,2}^n,...,R_{N_{S}-1,2}^n,R_{2,3}^n,...,R_{N_{S-1},3}^n,\cdots, R_{2,N_{L-1}}^n,R_{3,N_{L-1}}^n,...,R_{N_{S-1},N_{L-1}}^n)^{T},
\end{equation*}
\begin{equation*}
 \bf{b}=\begin{pmatrix}
 F_{2,2}V_{1,2}^{n+\frac{1}{2}} \\
 0 \\
 \vdots \\
  0\\
  E_{N_S-1, N_L-1}V_{N_S, N_L-1}^{n+\frac{1}{2}}
\end{pmatrix},
 \end{equation*}
 where
 \begin{align*}
	D_{i,j}&=1+\eta\Delta\tau k_0+\frac{r\eta\Delta\tau}{2}, \\
	E_{i,j}&=\frac{\eta\Delta\tau}{2}(k_0+(i-1)r), \\
	F_{i,j}&=\frac{\eta\Delta\tau}{2}(k_0-(i-1)r),  \\
	R_{i,j}^n&=\bigg[1-\Delta \tau(1-\eta)(k_0+\frac{r}{2})-\Delta \tau\bigg(\frac{\sigma_{L}^{2}}{{\Delta L}^{2}}+\frac{r}{2}\bigg)\bigg]V_{i,j}^{n}+\frac{(1-\eta)\Delta \tau}{2}\bigg[k_0+(i-1)r\bigg]V_{i+1,j}^{n}\\
	&+\frac{(1-\eta)\Delta \tau}{2}\bigg[k_0-(i-1)r\bigg]V_{i-1,j}^{n}+\frac{\Delta \tau}{2\Delta L}\bigg[\frac{\sigma_{L}^{2}}{\Delta L}+\alpha\bigg(\bar{\theta}+\lambda\kappa L_j^{\zeta}-L_j\bigg)\bigg]V_{i,j+1}^{n}\\
	&+\frac{\Delta \tau}{2\Delta L}\bigg[\frac{\sigma_{L}^{2}}{\Delta L}-\alpha\bigg(\bar{\theta}+\lambda\kappa L_j^{\zeta}-L_j\bigg)\bigg]V_{i,j-1}^{n}+k_1\bigg(V_{i+1,j+1}^{n}-V_{i+1,j-1}^{n}-V_{i-1,j+1}^{n}+V_{i-1,j-1}^{n}\bigg)\\
	&-\Delta \tau\sqrt{\frac{2}{\pi\delta t}}\kappa S_i\sqrt{k_2^2+k_3^2+k_4^2+2\rho_{1}k_2 k_3+2\rho_{2}k_3 k_4+2\rho_{3}k_2 k_4},
\end{align*}
with
\begin{align*}
	k_0&=(i-1)^{2}\bigg[\beta^{2}L_j^2+\sigma_{S}^{2}+2\rho_{1}\sigma_{S}\beta L_j\bigg],\\
	k_1&= \frac{(i-1)\Delta \tau}{4\Delta L}\bigg[\rho_{3}\sigma_{L}\beta L_j+\rho_{2}\sigma_{S}\sigma_{L}\bigg],\\
	k_2&= \frac{\beta(i-1)L_j}{\Delta S}\bigg[V_{i+1,j}^{n}-2V_{i,j}^{n}+V_{i-1,j}^{n}\bigg], \\
         k_3&= \frac{\sigma_{S}(i-1)}{\Delta S}\bigg[V_{i+1,j}^{n}-2V_{i,j}^{n}+V_{i-1,j}^{n}\bigg], \\
         k_4&= \frac{\sigma_{L}}{4\Delta S \Delta L}\bigg[V_{i+1,j+1}^{n}-V_{i+1,j-1}^{n}-V_{i-1,j+1}^{n}+V_{i-1,j-1}^{n}\bigg].
\end{align*}
For Eq. (\ref{step2}), the tridiagonal matrix $\mathcal{C}$ and vectors ${\bf{RRHS}}^{(n)}, \bf{c}$ are given as
\begin{equation*}
 \mathcal{C}=\begin{pmatrix}
 G_{2,2}&  -H_{2,2}& & & \\
  -M_{2,3}&  G_{2,3}  &&  & \\
  &  &\ddots& \ddots & \\
  &  &\ddots& \ddots &-H_{N_{S-1},N_{L-2}} \\
  &  & &-M_{N_{S-1},N_{L-1}} &  G_{N_{S-1},N_{L-1}}
\end{pmatrix}
\end{equation*}
\begin{equation*}
{\bf{RRHS}}^{(n)}_{k}=(RR_{2,2}^n,RR_{2,3}^n,...,RR_{2, N_L-1}^n,RR_{3,2}^n,...,RR_{3, N_L-1}^n,\cdots, RR_{N_S-1,2}^n,...,RR_{N_{S-1},N_{L-1}}^n)^{T},
\end{equation*}
\begin{equation*}
 \bf{c}=\begin{pmatrix}
 M_{2,2}V_{2,1}^{n+1} \\
 0 \\
 \vdots \\
  0\\
  H_{N_S-1, N_L-1}V_{N_S-1, N_L}^{n+1}
\end{pmatrix}
\end{equation*}
where
\begin{align*}
	G_{i,j}&=1+\eta\Delta\tau\bigg(\frac{\sigma_{L}^{2}}{{\Delta L}^2}+\frac{r}{2}\bigg), \\
	H_{i,j}&=\frac{\eta\Delta\tau}{2\Delta L}\bigg[\frac{\sigma_{L}^{2}}{\Delta L}+\alpha\bigg(\bar{\theta}+\lambda\kappa L_j^\zeta-L_j\bigg)\bigg],\\
	M_{i,j}&=\frac{\eta\Delta\tau}{2\Delta L}\bigg[\frac{\sigma_{L}^{2}}{\Delta L}-\alpha\bigg(\bar{\theta}+\lambda\kappa L_j^\zeta-L_j\bigg)\bigg],\\
	{RR}_{i,j}^n &= V_{i,j}^{n+\frac{1}{2}}+\bigg[\eta\Delta\tau\bigg(\frac{\sigma_{L}^{2}}{{\Delta L}^2}+\frac{r}{2}\bigg)\bigg]V_{i,j}^{n}-\frac{\eta\Delta\tau}{2\Delta L}\bigg[\frac{\sigma_L^2}{\Delta L}+\alpha\bigg(\bar{\theta}+\lambda\kappa L_j^\zeta-L_j\bigg)\bigg]V_{i,j+1}^{n}\\
          &-\frac{\eta\Delta\tau}{2\Delta L}\bigg[\frac{\sigma_L^2}{\Delta L}-\alpha\bigg(\bar{\theta}+\lambda\kappa L_j^\zeta-L_j\bigg)\bigg]V_{i,j-1}^{n},\\
\end{align*}

\section{EOC for the explicit finite difference scheme}

\begin{table}[H]
\caption{EOC in $\tau$-direction for $S_0=8$, $N_S=N_L=100$.}
\vskip5pt
\centering
\begin{tabular}{|c|c|c|c|c|c|c|}
\hline
\rowcolor[HTML]{EFEFEF}
\multicolumn{1}{|c|}{No. of steps} & \multicolumn{3}{c|}{Holder price} & \multicolumn{3}{c|}{Writer price}                      \\ \cline{2-7}
\rowcolor[HTML]{EFEFEF}
in $\tau$-direction  & Value     & Difference   & EOC    & Value     & Difference   & EOC \\ \hline
 250000  &  2.446939493 &  -           &  -           &  2.573456509 &  -           &  -             \\
 500000  &  2.446939185 &  3.07227E-07 &  -           &  2.573455789 &  7.19692E-07 &  -             \\
 750000  &  2.446939083 &  1.02293E-07 &  2.7123 &  2.573455507 &  2.82137E-07 &  2.3095   \\
 1000000 &  2.446939032 &  5.12766E-08 &  2.4006 &  2.573455391 &  1.16162E-07 &  3.0847   \\ \hline
\end{tabular}
\label{tauconvergence}
\end{table}

\begin{table}[H]
\caption{EOC in S-direction for $S_0=8$, $N_T=750000$, $N_L=150$.}
\vskip 5pt
\centering
\begin{tabular}{|c|c|c|c|c|c|c|}
\hline
\rowcolor[HTML]{EFEFEF}
\multicolumn{1}{|c|}{No. of steps} & \multicolumn{3}{c|}{Holder price} & \multicolumn{3}{c|}{Writer price}                      \\ \cline{2-7}
\rowcolor[HTML]{EFEFEF}
 in S-direction  & Value     & Difference   & EOC    & Value     & Difference   & EOC \\ \hline
5                                                                      & 2.5037                                          & -                                                                 & -                                             & 2.6193                                          & -                                                                 & -                                             \\
50                                                                      & 2.4365                                          & 0.0672                                                            & -                                             & 2.5663                                          & 0.0529                                                            & -                                             \\
75                                                                      & 2.4569                                          & 0.0204                                                            & 2.9353                                        & 2.5844                                          & 0.0181                                                            & 2.6481                                        \\
100                                                                     & 2.4469                                          & 0.01                                                              & 2.4960                                         & 2.5735                                          & 0.011                                                             & 1.7378                                        \\ \hline
\end{tabular}
\label{S_EOC_explicit}
\end{table}

\begin{table}[H]
\caption{EOC in L-direction for $S_0=8$, $N_T=750000$, $N_S=200$.}
\vskip 5pt
\centering
\begin{tabular}{|c|c|c|c|c|c|c|}
\hline
\rowcolor[HTML]{EFEFEF}
\multicolumn{1}{|c|}{No. of steps} & \multicolumn{3}{c|}{Holder price} & \multicolumn{3}{c|}{Writer price}                      \\ \cline{2-7}
\rowcolor[HTML]{EFEFEF}
 in L-direction  & Value     & Difference   & EOC    & Value     & Difference   & EOC \\ \hline
 25                                                                      & 2.44504367                                      & -                                                                 & -                                             & 2.57202047                                      & -                                                                 & -                                             \\
50                                                                      & 2.4456718                                       & 0.0006                                                            & -                                             & 2.572327033                                     & 0.0003                                                            & -                                             \\
75                                                                      & 2.445889855                                     & 0.0002                                                            & 2.6093                                   & 2.572525878                                     & 0.0002                                                            & 1.0677                                        \\
100                                                                     & 2.445792906                                     & 1E-04                                                             & 2.8176                                  & 2.572414049                                     & 0.0001                                                            & 2.0007                                        \\ \hline
\end{tabular}
\label{L_EOC_explicit}
\end{table}

Table 7-9 illustrates that the EOC is approximately order of two in both time and space directions.

\end{appendices}

\bibliographystyle{abbrv}
\bibliography{ref}

\end{document}